\documentclass[11pt,a4paper]{article}
\usepackage{jheppub}
\usepackage[english]{babel}
\usepackage{amssymb,amsmath}
\usepackage{mathrsfs}
\usepackage{hyperref}
\usepackage[dvipsnames,x11names]{xcolor}
\usepackage{graphicx}
\usepackage{subfig}
\usepackage{mathtools}
\usepackage{slashed}
\usepackage{float}
\usepackage{ulem}
\DeclarePairedDelimiter\abs{\lvert}{\rvert}%

\makeatletter
\gdef\@fpheader{}
\makeatother

\preprint{MIT-CTP/5829}

\title{Gauge hierarchy and metastability from Higgs-driven crunching}

\author[a]{Sean Benevedes,}
\author[b]{Ameen Ismail,}
\author[a,c]{and Thomas Steingasser}
 
\affiliation[a]{Department of Physics, Massachusetts Institute of Technology, Cambridge, MA 02139, USA}
\affiliation[b]{Enrico Fermi Institute, University of Chicago, Chicago, IL 60637, USA}
\affiliation[c]{Black Hole Initiative at Harvard University, 20 Garden Street, Cambridge, MA 02138, USA}

\emailAdd{seanmb@mit.edu}
\emailAdd{ameenismail@uchicago.edu}
\emailAdd{tstngssr@mit.edu}

\abstract{We present a new solution to the Higgs hierarchy problem based on dynamical vacuum selection in a landscape scanning the Higgs mass. In patches where the Higgs mass parameter takes a natural value, the Higgs potential only admits a minimum with a large and negative energy density. This causes a cosmological crunch, removing such patches from the landscape. Conversely, in patches where the Higgs mass parameter is smaller than a critical value, the Higgs potential admits a metastable minimum with the standard cosmological history. This critical value is determined by the instability scale, where the quartic coupling turns negative due to its running. The ability of this mechanism to explain the observed Higgs mass hinges on new physics at the TeV scale, such as vector-like fermions. We study two simple realizations of this scenario in a heavy neutral lepton model and in the singlet-doublet model, the latter mimicking a Higgsino-bino system. We show that the relevant parts of their parameter spaces can be probed by proposed future colliders, such as the FCC-ee or a muon collider.}

\begin{document}

\maketitle

\section{Introduction}
One of the most consequential open problems of modern fundamental physics is the apparent fine-tuning in the Standard Model (SM) Higgs potential. This includes the Higgs mass $M_H=125$~GeV and its vacuum expectation value (VEV) $v = 246$~GeV, giving rise to the \textit{(gauge) hierarchy problem}. As a relevant parameter unprotected by symmetry, the Higgs mass is sensitive to contributions from UV physics, so its natural value lies at the scale of new physics $\Lambda$ where the SM is embedded into some larger sector.
Conventional solutions to the hierarchy problem, based on protecting the Higgs mass with a symmetry, generically predict new states at the TeV scale. They are increasingly constrained by the lack of discovery of new physics at the Large Hadron Collider (LHC). At face value, the observed Higgs mass therefore appears severely fine-tuned, in contradiction with the principle of naturalness that has guided particle physics model building for decades~\cite{Buttazzo:2013uya,Giudice:2013yca,Giudice:2017pzm,Steingasser:2023ugv,Steingasser:2024hqi}.

This observation has galvanized the development of models in which the parameters of the low-energy Higgs potential are set through a dynamical mechanism in the early Universe~\cite{Graham:2015cka,Arkani-Hamed:2016rle,Geller:2018xvz,Cheung:2018xnu,Giudice:2019iwl,Strumia:2020bdy,Csaki:2020zqz,Arkani-Hamed:2020yna,Giudice:2021viw,TitoDAgnolo:2021nhd,TitoDAgnolo:2021pjo,Jung:2021cps,Domcke:2021yuz,Csaki:2022zbc,Chattopadhyay:2024rha}. These mechanisms often rely on the idea of vacuum selection, involving a landscape of vacua across which the Higgs mass parameter is scanned by UV physics, and a cosmological mechanism that selects patches in the landscape with a small Higgs VEV. The scanning could arise from, for example, a number of scalar fields that couple to the Higgs such that different minima of the potential along the direction of these scalars give rise to different Higgs parameters. The selection mechanism then selects vacua by, for instance, causing certain patches to undergo a cosmological crunch.\footnote{See also Refs.~\cite{Bloch:2019bvc,Csaki:2024ywk} for applications of this idea to the cosmological constant problem and the doublet-triplet splitting problem.}

Recent works have suggested a connection between the hierarchy problem and the metastability of the electroweak (EW) vacuum. Extrapolating from the observed Higgs and top quark mass, it appears that (within the SM) the Higgs quartic coupling turns negative above the \textit{instability scale} $\mu_I \sim 10^{10}$~GeV~\cite{Degrassi:2012ry,Buttazzo:2013uya,Espinosa:2015qea}, implying the EW vacuum is metastable. This metastability is of significance for the mechanism developed in Refs.~\cite{Khoury:2019ajl,Khoury:2019yoo,Kartvelishvili:2020thd,Khoury:2021grg}, which predicts metastable vacua with a specific lifetime. In contrast to the mechanisms developed e.g.\ in Refs.~\cite{Graham:2015cka,Arkani-Hamed:2016rle,Geller:2018xvz,Cheung:2018xnu,Giudice:2019iwl,Strumia:2020bdy,Csaki:2020zqz,Arkani-Hamed:2020yna,Giudice:2021viw,TitoDAgnolo:2021nhd,TitoDAgnolo:2021pjo,Jung:2021cps,Domcke:2021yuz,Csaki:2022zbc,Chattopadhyay:2024rha}, this analysis does not rely on a concrete mechanism in the early universe, but rather a statistical analysis of the landscape as a whole. These results have been extended to the hierarchy problem through the \textit{metastability bound} on the Higgs mass in Refs.~\cite{Khoury:2021zao,Benevedes:2024tdq}. This bound states that requiring EW vacuum metastability implies an unnaturally small Higgs mass under very general conditions --- independently of the landscape picture of Refs.~\cite{Khoury:2019ajl,Khoury:2019yoo,Kartvelishvili:2020thd,Khoury:2021grg}.

In this article, we introduce a new approach to dynamical vacuum selection using Higgs metastability. Similar to existing works, we postulate the existence of a UV completion which scans the Higgs mass over the landscape, and furthermore demand it to give rise to a Higgs potential admitting one true vacuum with a large and negative energy density. Patches with generic Higgs parameters undergo a crunch once the Higgs field settles into this vacuum. However, in patches with an unnaturally small Higgs mass and VEV, an additional, metastable vacuum forms near the local maximum at $h = 0$ through renormalization group (RG) running. In the false vacuum the standard cosmological history unfolds. Hence the only patches that survive until the present day are those in which the Higgs mass appears fine-tuned. Since our mechanism is not probabilistic, our main conclusion --- that this picture requires vacua with an unnaturally small Higgs mass --- does \textit{not} directly rely on the choice of a measure on the landscape.

In section~\ref{sec:mechanism} we discuss the conditions for the formation of a metastable vacuum in the Higgs potential. Our central observation is that, from the perspective of the potential of the UV theory, the metastable vacuum can be understood as being formed ``on top'' of a local maximum through quantum corrections to the quartic coupling, allowing it to avoid crunching. We find that a metastable minimum compatible with the spontaneous symmetry breaking (SSB) pattern of the SM can only form if the Higgs mass parameter $m^2$ is less than a critical value $m_{\rm crit}^2$. In Sec.~\ref{subsec:m2gtr0}, we review a concrete example for a vacuum selection mechanism capable of eliminating vacua without this pattern to which our argument applies. This critical value is set by the instability scale $\mu_I$ at which the Higgs quartic turns negative. One can have a naturally large hierarchy between $\mu_I$ and the UV cutoff $\Lambda$ since the former is generated by the running of the quartic coupling. We find that the difference in energy density between the EW vacuum and the true vacuum (associated with a natural Higgs VEV) also depends on this hierarchy: a large hierarchy leads to a higher relative energy density of the EW vacuum.

The critical value $m_{\rm crit}$ is proportional to the instability scale $\mu_I$, but $\mu_I \gg 1$~TeV in the SM~\cite{Steingasser:2023ugv,Detering:2024vxs}. Therefore, explaining the observed value of the EW scale in our mechanism requires new physics capable of lowering $\mu_I$. In section~\ref{sec:models} we study the phenomenology of two simple models that achieve this by introducing vector-like leptons. The first involves just a singlet fermion, which we find can lower the Higgs mass bound to $\sim 2$~TeV while remaining consistent with experimental constraints. The second is the well-known singlet-doublet model, which is more weakly constrained and thus allows for a bound as strong as $\sim 700$~GeV. These bounds can be expected to be too conservative, as saturation would correspond to a potential with no barrier protecting the EW vacuum. The relevant parameter space of both models can be probed by future lepton colliders including the ILC, the FCC-ee, and a muon collider. 

Lowering the instability scale in this way also reduces the lifetime of the metastable electroweak vacuum. Our mechanism needs to be supplemented with some new physics that partially stabilizes the vacuum, so as to ensure that our patch should not have already decayed. We comment on this in section~\ref{sec:cosmology}, 
where we provide an estimate of the scale at which higher-dimensional operators representing UV physics need to have a significant stabilizing effect on the tunneling rate. We summarize our findings in section~\ref{sec:conclusions}.

Our work demonstrates that \textit{any} scanning mechanism satisfying two simple conditions automatically offers an explanation for the hierarchy problem, which is testable through its dependence on new physics at the TeV scale. Complementary to this, a series of recent works have founded a rich phenomenological program centered around ways of lowering $\mu_I$. This pattern has been independently proposed in Refs.~\cite{Khoury:2021zao,Giudice:2021viw} (using a relation first observed in Ref.~\cite{Buttazzo:2013uya}), which established its viability within current experimental constraints for a vector-like fermion and right-handed neutrinos, respectively. In the context of a similar bound for a positive mass term, Ref.~\cite{Benevedes:2024tdq} demonstrated that the subset of right-handed neutrino parameter space giving rise to relevant bounds could be probed by realistic next-generation colliders. More recently, it was suggested in Refs.~\cite{Enguita:2025ybx,Steingasser:talk} to consider ``complete models'' which naturally contain additional stabilizing physics at scales just above the instability scale, such as the Majoron model. An alternative approach has been brought forward in Refs.~\cite{Detering:2024vxs,Detering:2025nmh}, whose authors demonstrate that the instability scale could also be lowered through the effects of an axion-like particle lighter than the Higgs through a nontrivial mixing.

\section{Higgs mass and vacuum energy density}\label{sec:mechanism}
We consider a scenario in which the Higgs sector is embedded into an extension with a vacuum structure suitable for scanning the Higgs mass parameter and the cosmological constant. In the following, we will argue that \textit{any} scanning mechanism subject to two conditions inevitably gives rise to a parametrically large hierarchy between the Higgs mass and its natural value $m^2\sim \Lambda^2 $.

\subsection{Basic idea}

First, we require the scanning sector to only allow for vacua giving rise to the EW symmetry breaking pattern, i.e., a negative effective Higgs mass parameter near all vacua of interest. In Sec.~\ref{subsec:m2gtr0} we present an example for a well-established scanning mechanism capable of achieving this and comment on the realization of our argument in this model.


Second, we assume that all \textit{natural vacua} of the theory (meaning VEVs and masses of order $\Lambda$) have a negative energy density.
This condition could, for example, arise from the form of the underlying theory's potential, and relates to the famous de Sitter swampland conjecture~\cite{Obied:2018sgi,Garg:2018reu,Ooguri:2018wrx}. Any Hubble patch where the Higgs rolls down to such a vacuum will undergo a cosmological crunch due to the negative vacuum energy, removing that patch from the landscape. Starting from this premise, addressing the hierarchy problem amounts to showing that an unnaturally small Higgs mass allows for the formation of a metastable vacuum with an energy density elevated above the natural vacua. This metastable vacuum does not crunch and can be cosmologically long-lived. Crucially, this circumvents the measure problem underlying many vacuum selection mechanisms.\footnote{In the language of Refs.~\cite{TitoDAgnolo:2021nhd,TitoDAgnolo:2021pjo}, this amounts to a dynamical selection as opposed to a statistical selection.}

\begin{figure}[t]
    \centering
     \raisebox{-0.5\height}{\includegraphics[width=0.425\textwidth]{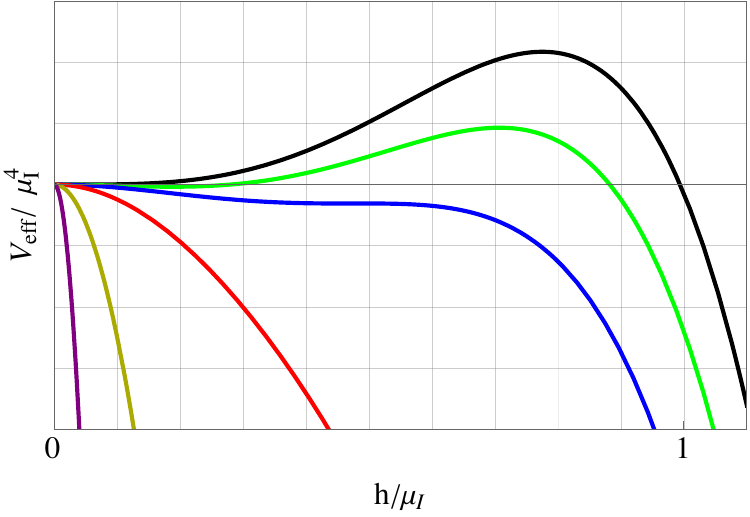}} \raisebox{-0.5\height}{\includegraphics[width=0.41\textwidth]{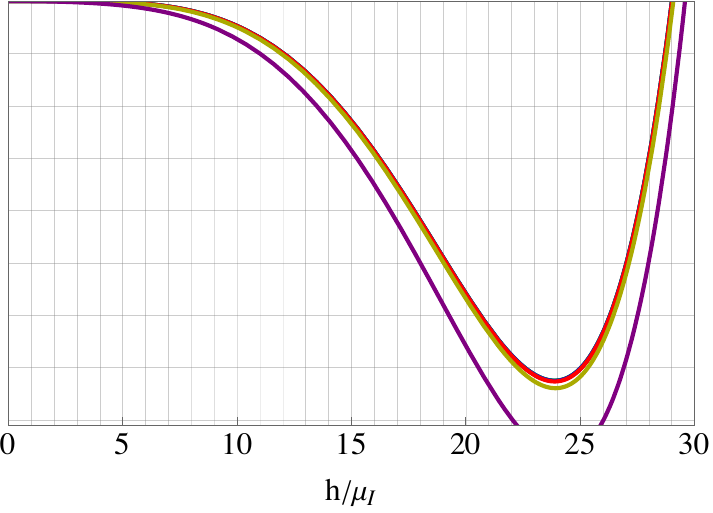}}\raisebox{-0.4\height}{\includegraphics[width=0.135\textwidth]{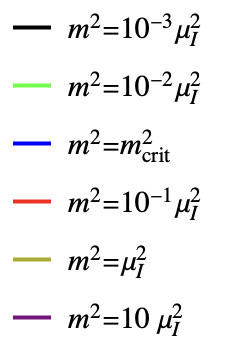}} 
    \caption{A sketch of the effective Higgs potential with $\lambda (\mu = \Lambda) <0$ stabilized by a higher-dimensional operator $\Delta V = h^6/\Lambda^2$. We note that while this effect implies a strong dependence of the potential around the EW vacuum on the scanned parameter $m^2$ (left panel), while the behavior of the potential around the natural minimum is mostly independent of $m^2$. For values of the mass parameter smaller than some critical value $m_{\rm crit}^2\sim \mu_I^2$ (see Eq.~\eqref{eq:mcrit}), a second metastable minimum can form. Due to the logarithmic running of this coupling, this happens at field values several orders of magnitude smaller than $\Lambda$. Here, we have for concreteness chosen $\beta_\lambda (\mu_I)=-0.1$ and $\Lambda=100 \mu_I$. 
    }
    \label{fig:potential_sketch}
\end{figure}

From a bottom-up perspective, this assumption is motivated by the structure of the Standard Model Higgs potential, which is generally of the form
\begin{align}\label{eq:VeffLE}
    V_{\rm eff}(h) = - \frac{m^2}{4} h^2 + \frac{\lambda}{4} h^4 + \sum_{n=1}^\infty \frac{C_{2n+4}}{\Lambda^{2n}}  h^{2n+4} + V_0,
\end{align}
where the constant term $V_0$ includes gravitational contributions to the cosmological constant. All current measurements suggest that for large energies both $m^2>0$ and $\lambda <0$, also implying the presence of a minimum with a large negative energy density, satisfying our condition. In the following, we will find that such a vacuum is indeed a generic feature of this potential, contrasting the EW vacuum, whose existence relies on some degree of fine-tuning. We illustrate this in Fig.~\ref{fig:potential_sketch} for the toy-model we consider in the rest of this section. The requirement for this second minimum to \textit{always} crunch is therefore equivalent to an upper bound on the possible values of the cosmological constant achievable by the scanning mechanism.

In the following subsections, we will argue that in patches with a small mass term radiative effects can lead to the formation of an additional vacuum with a significantly higher energy density. As stated before, our mechanism is independent of the details of the scanning mechanism beyond the previously laid out conditions. Instead, the ability of this argument to explain the observed Higgs mass hinges on physics near the EW scale that is, a priori, unrelated to the scanning mechanism. In Sec.~\ref{sec:models}, we will discuss prospects of probing this new physics in future experiments.

\subsection{Tree-level estimates}
\label{sec:tree}
Extrapolating the SM to sufficiently high energy scales, the existence of at least one additional, lower-lying vacuum can be deduced from the running of the quartic coupling $\lambda$, which plays a crucial role in the metastability bound on the Higgs mass. To illustrate the importance of the running for our argument, we will first analyze the vacuum structure of the effective potential, Eq.~\eqref{eq:VeffLE}, at tree-level. This will lead us to conclude that running effects on $\lambda$ need to be significant for our mechanism to be captured in the low-energy EFT.

For the tree-level potential, it is straightforward to obtain approximate relations for all of its extrema and their respective energy densities. We will assume a generic set of Wilson coefficients and first consider the case $C_6 >0$. This makes it easy to see that the potential Eq.~\eqref{eq:VeffLE} has a unique minimum at some $H=v_H$. At leading order in $m^2/\Lambda^2$ it is given by
\begin{alignat}{2}
    v^2 &\simeq  \frac{|\lambda | \Lambda^2}{6 C_6} , \ \ &&V_{\rm eff}(v) \simeq - |\lambda|^3 \frac{\Lambda^4}{432 C_6^2} +V_0  \ \ \ \text{for} \ \lambda < 0 , \ \text{while} \label{eq:vH-} \\ 
    v^2 &\simeq \frac{m^2}{2 \lambda} , \ \ &&V_{\rm eff}(v) \simeq - \frac{m^4}{16 \lambda}+V_0 \ \ \ \text{for} \ \lambda > 0 . \label{eq:vH+}
\end{alignat}
Note that the latter case agrees to leading order with the result from including renormalizable operators alone. The first of these vacua represents one of our previously discussed natural vacua, with energy $V_{\rm eff}(v) \sim -\Lambda^4$, and is sensitive to higher-dimensional operators in the potential. In these terms, our assumption regarding the negativity of the vacuum energy would translate to the inequality $V_0 < |\lambda|^3 \frac{\Lambda^4}{432 C_6^2}$.

Thus, in a landscape satisfying our previous assumptions, the existence of a non-crunching vacuum requires $\lambda >0$ near the vacuum. For a natural mass parameter $m^2 \sim \Lambda^2$, this scenario would also give rise to a natural vacuum, already suggesting the need for a small mass parameter. While this observation is consistent with our own vacuum, it can not be linked to any further phenomenological consequences. In particular, this includes the possibility to relate the properties of our vacuum to the landscape as a whole: for a strictly positive $\lambda$, the natural vacuum closest to the electroweak one can only form through the interplay of at least two higher-dimensional operators (e.g.\ if $C_6 <0$ and $C_8>0$). Hence, the properties of this vacuum, including its energy, depend predominantly on the details of these unknown operators. Moreover, such a scenario is not necessarily generic from a top-down perspective. Negative threshold corrections to the quartic coupling are a common feature of extensions of the SM by additional (pseudo-)scalars, which can be expected to play a crucial role in a landscape~\cite{Jiang:2018pbd}. 

However, this picture is incomplete: the condition $\lambda >0$ is famously scale-dependent. In the RG-corrected effective potential this manifests as a dependence of $\lambda$ on the field value.
As mentioned in the introduction, this is indeed the case in the SM, where the quartic turns negative at $\mu_I \sim 10^{10}$~GeV. This suggests to take seriously the possibility that $\lambda >0$ for small field values $H \ll \Lambda$, leading to a minimum described by Eq.~\eqref{eq:vH+}, while $\lambda <0$ near $H\sim \Lambda$, giving rise to a second minimum qualitatively described by Eq.~\eqref{eq:vH-}. Hence, taking into account the running of $\lambda$ allows for a third scenario which combines the two existing at tree-level. In section~\ref{sec:OLE} we discuss the effect of the running of $\lambda$ on our argument, before extending it in section~\ref{sec:OLA} by considering the interplay with higher-dimensional operators.

\subsection{One-loop estimates}
\label{sec:OLE}
We can approximate the effective potential to one-loop accuracy by
\begin{align}\label{eq:Vefffull}
    V_{\rm eff} (h,\mu=h) = V_0 - \frac{m^2}{4} h^2 +  \frac{\beta_{\lambda}}{4} \ln \left( \frac{h}{\mu_I} \right) \cdot h^4 + ...,
\end{align}
where the instability scale $\mu_I$ is defined by $\lambda_{\rm eff} (h=\mu_I,\mu=\mu_I)=0$, including non-logarithmic one-loop corrections. For now we neglect all higher-dimensional operators so we can make simple analytical estimates. These are, however, only reliable in the limit of a large hierarchy $\mu_I / \Lambda \ll 1$. We will provide a more complete discussion in section~\ref{sec:OLA}, taking into account the effects of higher-dimensional operators.

The vacuum structure of this potential has previously been studied in great detail in Refs.~\cite{Buttazzo:2013uya,Khoury:2021zao,Steingasser:2023ugv,Steingasser:2024hqi}, revealing the existence of two distinct \textit{quantum phases}, assuming $\lambda <0$ at the matching scale $\Lambda$. The transition between the phases occurs at the critical value of $m^2$,
\begin{align}\label{eq:mcrit}
    m_{\rm crit}^2 = e^{-3/2} |\beta_\lambda (\mu_I)| \mu_I^2.
\end{align}
For $m^2 < m_{\rm crit}^2$, we recover a similar behavior to that of the SM, where a metastable vacuum described by Eq.~\eqref{eq:vH-} forms. For larger values the potential barrier disappears, such that the only possible vacua are ones that form through the effects of higher-dimensional operators, as described by Eq.~\eqref{eq:vH+}. As pointed out below that equation, these give rise to a natural Higgs VEV, and hence crunch by our assumption. See Fig.~\ref{fig:potential_sketch}. Eq.~\eqref{eq:mcrit} establishes the metastability bound~\cite{Khoury:2021zao}.

Strictly speaking, we are interested in obtaining a bound on the physical Higgs mass $M_H^2$, as opposed to the mass parameter. We study the physical Higgs mass in appendix~\ref{sec:stepfunction} and find that $M_H^2 \leq e^{-1} m_{\rm crit}^2$. Going beyond the simple approximation of a constant $\beta$ function would change the precise bound, but we still expect $M_H^2$ to be bounded from above by $\mathcal{O}(m_{\rm crit}^2)$.


Fig.~\ref{fig:potential_sketch} also illustrates another crucial feature of this scenario: generically, the resulting minimum is formed ``on top'' of a local maximum of the full potential evaluated at this RG scale. In the context of our landscape picture, this implies that these \textit{radiatively generated vacua} (RGV) generically have a higher vacuum energy density than their natural counterparts. From this perspective, the metastability bound can be understood as a necessary condition for the formation of such a vacuum. If it is violated, only vacua with large and negative energy densities form, all of which would quickly crunch away. Moreover, in this picture the negativity of $\lambda$ near the matching scale $\Lambda$ is a typical feature of the local maximum ``on top'' of which this new minimum is formed.

Following our discussion in section~\ref{sec:tree}, and in particular Eqs.~\eqref{eq:vH-} and~\eqref{eq:vH+}, we are able to estimate the elevation of the EW energy density relative to its adjacent natural vacuum. However, Eqs.~\eqref{eq:vH-} and~\eqref{eq:vH+} do not take into account the running of $\lambda$, making them unreliable when $m^2 \sim m_{\rm crit}^2$. Likewise, as previously mentioned, we expect the dynamics to be sensitive to higher-dimensional operators for relatively small hierarchies between $\mu_I$ and $\Lambda$.
Thus, we provide an extended version of these bounds in the next subsection, taking into account the running of $\lambda$ as well as the interplay between the instability scale and higher-dimensional operators. Although this is necessary to obtain estimates for the energy density of the RGV, we will find that the impact of these operators on the metastability bound Eq.~\eqref{eq:mcrit} is~$\mathcal{O}(10\%)$, making it negligible for our phenomenological analysis in section~\ref{sec:models}.

\subsection{Full analysis}
\label{sec:OLA}
\begin{figure}[h]
    \centering
    \includegraphics[width=0.8\textwidth]{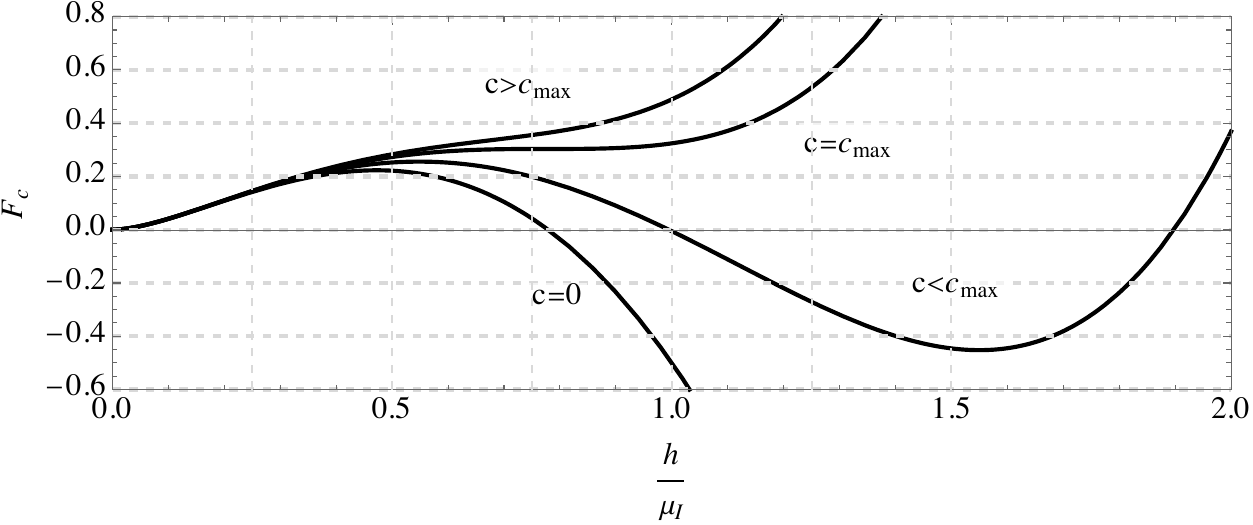}
    \caption{A visualization of Eq.~\eqref{eq:criteqfull}. The curves represent the function $F$ for different values of the parameter $c = C_6 \mu_I^2/(\Lambda^2 |\beta_\lambda (\mu_I)|)$, which encodes the corrections from higher-dimensional operators. For small enough or vanishing values of this parameter, the function $F$ has a local maximum. If $m^2$ is larger than this maximum, the function $F$ intersects the $y=m^2$-line once, corresponding to the quantum phase of a unique minimum. For smaller values of $m^2$, the three intersections correspond to an additional minimum with adjacent potential barrier. For large enough values of $c$, corresponding to a small UV scale $\Lambda$, there always exists a unique minimum, independent of $m^2$. In this regime, the dimension-six operator becomes important already in the regime where $\lambda >0$.}
    \label{fig:F}
\end{figure}

Keeping the leading-order higher-dimensional term in Eq.~\eqref{eq:Vefffull} (i.e.\ $C_6 h^6 / \Lambda^2$), the equation for a local extremum can be brought to the form
\begin{align}\label{eq:criteqfull}
    \frac{m^2}{|\beta_\lambda (\mu_I)| \mu_I^2}= 2 \frac{h^2}{\mu_I^2}\left[ 6 \frac{C_6}{ |\beta_\lambda (\mu_I)|} \frac{\mu_I^2}{ \Lambda^2} \frac{ h^2}{\mu_I^2}  - \ln \left( \frac{h}{\mu_I} e^{1/4}\right) \right] \equiv F_c(h/\mu_I) ,
\end{align}
where we define the auxiliary parameter $c= C_6 \mu_I^2/(\Lambda^2 |\beta_\lambda (\mu_I)|)$. The different phases of the potential are now distinguished by the number of solutions to this equation for a given $m^2$. Eq.~\eqref{eq:criteqfull} suggests the simple geometric interpretation shown in Fig.~\ref{fig:F}. 

\begin{figure}[t!]
    \centering
    \includegraphics[width=0.48\textwidth]{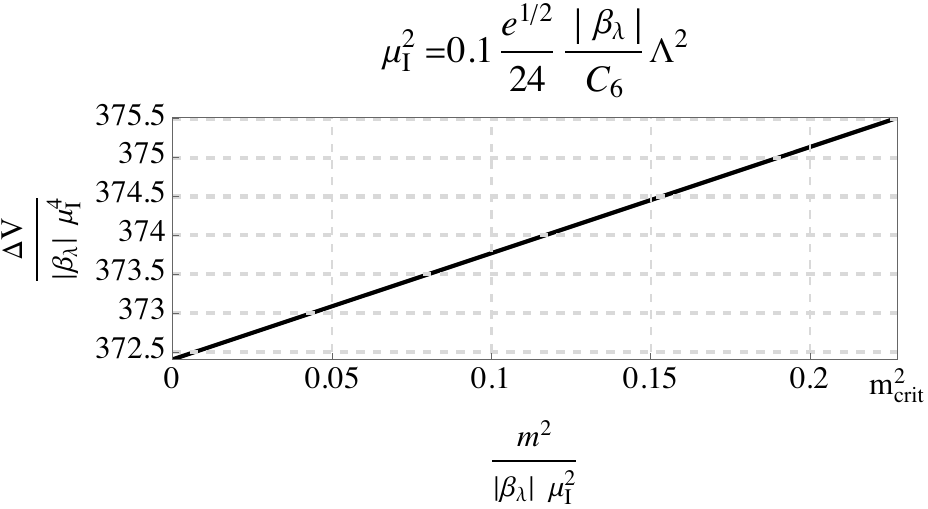} \includegraphics[width=0.48\textwidth]{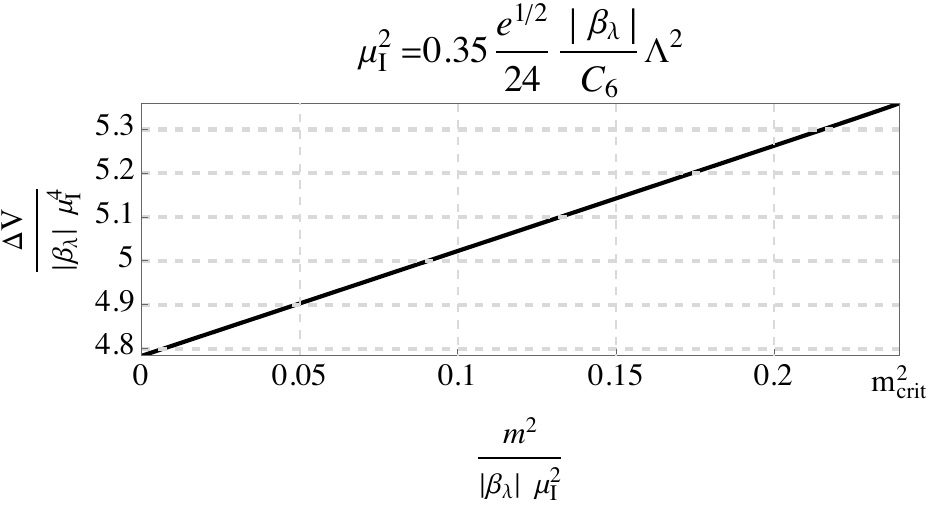}
    \includegraphics[width=0.48\textwidth]{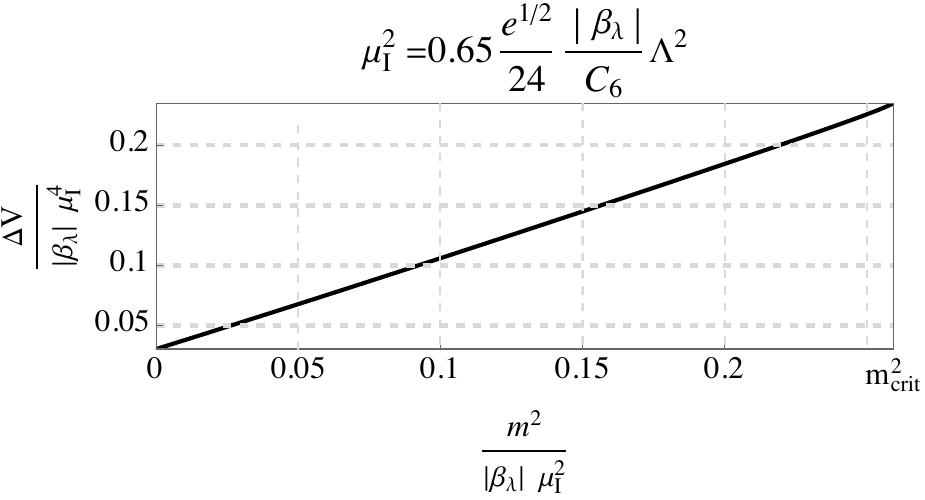}
    \includegraphics[width=0.48\textwidth]{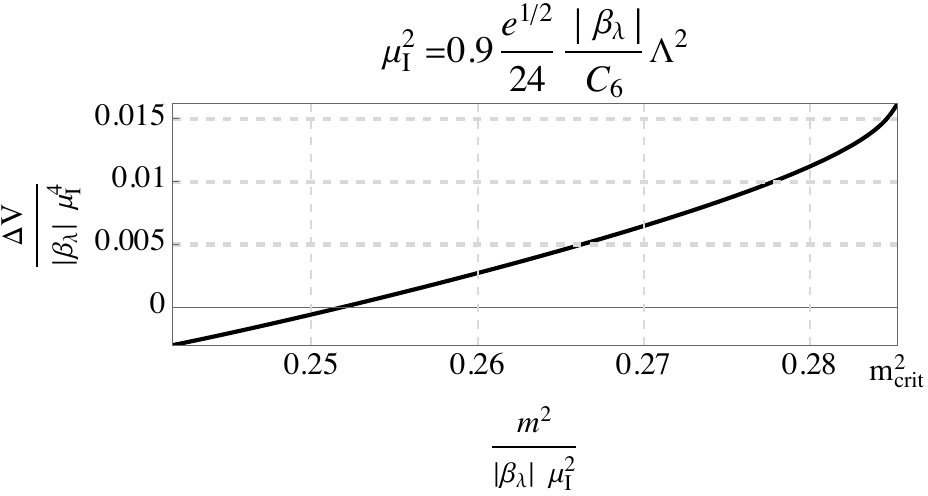}
    \caption{The elevation of the radiatively generated vacuum as a function of $m^2$ in units of $\mu_I^4 |\beta_\lambda (\mu_I)|$ and $\mu_I^2  |\beta_\lambda (\mu_I)|$, respectively, for different hierarchies between the instability scale and the scale of new physics. }
    \label{fig:dV}
\end{figure}

First consider the case $c>c_{\rm max}\equiv \sqrt{e}/24$, corresponding to a single solution of Eq.~\eqref{eq:criteqfull}. This case mimics the deep vacuum described by Eq.~\eqref{eq:vH+} in Sec.~\ref{sec:tree}, which always crunches under our assumptions. In what follows we therefore focus on the scenario $c <c_{\rm max}$. For these cases, the function $F_c(h)$ has one local maximum at $h_{\rm max}$. Thus, if $m^2 > |\beta_\lambda (\mu_I)| F_c (h_{\rm max}) \mu_I^2$, Eq.~\eqref{eq:criteqfull} has a single solution, representing the natural vacuum. This immediately allows us to identify the full version of Eq.~\eqref{eq:mcrit}, taking into account the corrections from the dimension-six term:
\begin{align}
    m^2<m_{\rm crit}^2 = |\beta_\lambda (\mu_I)| F_c (h_{\rm max}) \mu_I^2 .  
\end{align}
In addition, the condition that $c<c_{\rm max}$ represents a lower bound on the scale $\Lambda$ in terms of the instability scale,
\begin{align}
     \Lambda^2 > \frac{24 C_6}{\sqrt{e} |\beta_\lambda (\mu_I)|} \mu_I^2 
\end{align}
When this inequality is satisfied, it is straightforward to calculate the energies of the two vacua and compare their energies. We present our results in Fig.~\ref{fig:dV}. For notational clarity, we will from now on denote the difference in energy density between the RGV and its adjacent natural vacuum as its \textit{elevation}.

\begin{figure}[t!]
    \centering
    \includegraphics[width=0.8\textwidth]{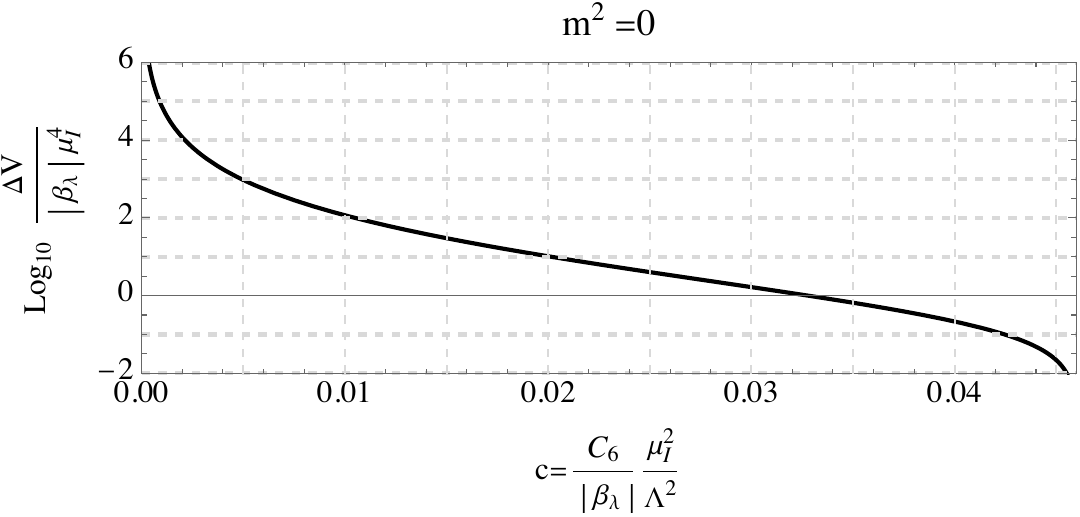} 
    \caption{The elevation of the radiatively generated vacuum relative to its adjacent natural vacuum in units of $\mu_I^4/|\beta_\lambda (\mu_I)|$ as a function of the parameter $c$ for $m^2=0$. For $0.046 \leq c < c_{\rm max}$, the second minimum still exists, but has a larger energy density than its counterpart at $h=0$.}
    \label{fig:f}
\end{figure}

First, we note that the elevation of the RGV is mostly independent of the mass parameter as long as $c / c_{\rm max} \lesssim 1/3 $. That is, 
\begin{align}
    \mu_I^2 \lesssim \frac{1}{3} \frac{\sqrt{e}}{24} \frac{|\beta_\lambda (\mu_I)|}{C_6} \Lambda^2 \Leftrightarrow \text{negligible}\ m^2\text{-dependence}.
\end{align}
For the dependence on $c$, we find that smaller values --- i.e., a larger hierarchy between $\mu_I$ and $\Lambda$ --- correspond to a larger elevation in units of $\mu_I^4$. This dependence of the elevation on $c$ is presented in Fig.~\ref{fig:f}.

Next, if $1/3 \lesssim c/ c_{\rm max} \lesssim 2/3 $, we also find that the elevation depends predominantly on the value of $c$. However, unlike the previous case, we find that the elevation shows a subdominant but relevant dependence on the mass parameter, with masses closer to the critical value corresponding to larger elevations, 
\begin{align}
    \frac{1}{3} \frac{\sqrt{e}}{24} \frac{|\beta_\lambda (\mu_I)|}{C_6} \Lambda^2 \lesssim \mu_I^2 \lesssim \frac{2}{3} \frac{\sqrt{e}}{24} \frac{|\beta_\lambda (\mu_I)|}{C_6} \Lambda^2 \Leftrightarrow \text{moderate}\ m^2\text{-dependence}.
\end{align}
In particular, for $c / c_{\rm max} \simeq 2/3$ and in the limit of a vanishing Higgs mass, the two vacua can become degenerate.

Lastly, for $2/3 \lesssim c / c_{\rm max} < 1$, the dependence on $m^2$ is enhanced such that the RGV can lie below the natural vacuum. If this is the case, then the RGV would crunch just like the natural vacuum. More generally, we observe in agreement with our previous discussion that lowering the hierarchy between $\mu_I$ and $\Lambda$ to this extent leads to a small elevation even for near-critical masses.

Altogether, this implies that the existence of an RGV which is significantly elevated above the natural vacuum requires
\begin{align}\label{eq:totalhierarchy}
    m^2 \lesssim e^{-3/2} |\beta_\lambda (\mu_I)| \mu_I^2 \lesssim f(\Delta V) \frac{e^{-1/2}}{24 C_6} |\beta_\lambda (\mu_I)|^2 \Lambda^2.
\end{align}
Here, $0<f(\Delta V)<1$ controls the elevation of the EW minimum to leading order, where smaller values correspond to a larger elevation.

\subsection{Positive $m^2$}
\label{subsec:m2gtr0}
So far we have restricted our discussion to vacua with negative Higgs mass parameter. Our mechanism crunches away patches with a large Higgs VEV, but has no effect on patches where the Higgs mass parameter is positive and the VEV vanishes. We have not fully addressed the hierarchy problem until we also introduce a mechanism for excising patches with positive $m^2$ from the landscape. Here we sketch a straightforward way to accomplish this.

There exist dynamical mechanisms in the literature which cause patches with positive Higgs mass parameter to undergo a crunch, e.g. Refs.~\cite{Strumia:2020bdy,Csaki:2020zqz,TitoDAgnolo:2021nhd,TitoDAgnolo:2021pjo}. It is easy to realize one of these mechanisms within our scenario. As a concrete example, following Ref.~\cite{TitoDAgnolo:2021nhd} we can introduce an ultralight scalar $\phi$ with an approximate shift symmetry, broken by the potential
\begin{equation}
    V_\phi = -\frac{1}{2} m_\phi^2 \phi^2 - \frac{1}{4} \lambda_\phi \phi^4 - \frac{\alpha_s}{8\pi f} \phi G \tilde{G} .
\end{equation}
The scalar mass is technically natural since it breaks the shift symmetry, and we expect $\lambda_\phi \sim m_\phi^2 / M^2 \ll 1$, where $M$ is the scale at which the shift symmetry is broken. We also include an axion-like coupling to the the gluon, which is generated at a UV scale $f$.

Below the QCD scale, nonperturbative corrections generate a term in the potential of the form $-\Lambda^4 \cos(\phi/f)$. The scale $\Lambda^4 \sim f_\pi^2 m_\pi^2 \propto v$ grows linearly with the Higgs VEV in different patches~\cite{Graham:2015cka}. Expanding the potential about the origin then gives
\begin{equation}
    V_\phi = \frac{1}{2} \left( -m_\phi^2 + \frac{\Lambda^4}{f^2} \right) \phi^2 - \frac{1}{4} \left(\lambda + \frac{\Lambda^4}{f^4}\right) \phi^4 + \dots
\end{equation}
Hence a stable minimum for $\phi$ only exists if $\Lambda^4 \gtrsim m_\phi^2 f^2$. Since $\Lambda^4$ grows with $v$, in patches with a small or vanishing Higgs VEV $\phi$ does not have a stable minimum; it rolls down and triggers a crunch.

The phenomenology of this mechanism was discussed at length in Ref.~\cite{TitoDAgnolo:2021nhd} and is essentially disconnected from that of our scenario. This mechanism serves as a proof of concept, but there likely exists a plethora of models for the scanning sector excluding vacua with a positive mass parameter, compatible with the scenario discussed in this work.


\section{Minimal models}\label{sec:models}
We now explore simple models that allow us to address the hierarchy problem within the context of our mechanism. As a reminder, we assume a landscape of causally disconnected Hubble patches, where a scanning sector sets the value of the Higgs mass in each patch. We have seen that patches in the multiverse undergo a crunch unless the Higgs mass is less than the critical value $m_{\rm crit}$, which scales linearly with the instability scale $\mu_I$. Hence the only patches which survive until today are those in which the Higgs mass is smaller than $m_{\rm crit}$, leading to an apparent fine-tuning.

However, the instability scale $\mu_I$ in the SM is many orders of magnitude above the EW scale. To dynamically select the electroweak scale, we need to introduce new fermions to push down $\mu_I$ to $\mathcal{O}({\rm TeV})$. We focus here on two simple models where the new fermions are $SU(3)$ color singlets.
The first involves a heavy neutral lepton mixing with the third lepton generation (it could also mix with the other generations, but this is more constrained experimentally), and the second is the singlet-doublet model. Similar ideas were suggested in Refs.~\cite{Giudice:2021viw,Khoury:2021zao} as a means of lowering the instability scale. Refs.~\cite{Enguita:2025ybx,Steingasser:talk,Detering:2025nmh} have recently presented the idea of testing this mechanism in future colliders such as FCC using a model involving right-handed neutrinos, which we reproduce.

Our reason for studying $SU(3)$ singlet fermions is that the resulting phenomenology is distinct from symmetry-based solutions to the hierarchy problem, which generically include TeV-scale top partners. These do not appear in the models we consider. Note that although the new fermions appear near the TeV scale, they have nothing to do with canceling large contributions to the Higgs mass.

These models are also different from other approaches to Higgs naturalness based on dynamical vacuum selection. Those generically predict a light scalar which weakly mixes with the Higgs. The potential for the scalar is sensitive to TeV-scale Higgs VEVs, which allows the selection of a realistic electroweak scale. In contrast, our mechanism does not invoke light scalars, as the Higgs potential itself is sensitive to a TeV-scale Higgs VEV due to the introduction of the new fermions.
 
\subsection{Heavy neutral lepton}
The simplest possibility is to introduce a pair of vector-like SM singlets ($SU(2)$ singlet, no hypercharge) $\psi_{L,R}$. The Lagrangian is
\begin{equation}
    \mathcal{L}_{\psi} = i \overline{\psi_L} \slashed{\partial} \psi_L + i \overline{\psi_R} \slashed{\partial} \psi_R - \left( M \overline{\psi_L} \psi_R + y_i \overline{L_i} \tilde{H} \psi_R + {\rm ~h.c.} \right)
\end{equation}
The model has four free parameters, the fermion mass $M$ and the Yukawa couplings to the SM $y_i$ (one for each generation). If we were to also add a lepton number-violating Majorana mass term for $\psi$, we would generate neutrino masses through the inverse seesaw mechanism~\cite{Mohapatra:1986bd,Gonzalez-Garcia:1988okv}. Here we are not interested in generating neutrino masses, though, so we omit this term.

For simplicity and concreteness, let us assume that the new fermion only couples to one of the lepton doublets. Specifically we will take it to couple to the third generation, so $y_\tau = y$ and $y_{e,\mu} = 0$, because this choice will give the least stringent experimental bounds. After EW symmetry breaking, the neutrino $\nu_\tau$ mixes with $\psi$ through an angle
\begin{equation}\label{eq:mixingangle}
    \sin \theta = \frac{yv/\sqrt{2}}{\sqrt{y^2 v^2/2 + M^2}} .
\end{equation}
In the heavy neutral lepton literature $\sin^2 \theta$ is usually referred to as $\abs{V_{\tau N}}^2$, $\abs{U_{\tau 4}}^2$, or similar~\cite{Abdullahi:2022jlv,CMS:2024xdq}.

This fermion modifies the running of the Higgs quartic coupling $\lambda$ above the scale $M$ through two additional terms in its $\beta$ function,
\begin{equation}\label{eq:onelooprg}
    \Delta \beta = \frac{1}{16\pi^2} \left[ -2 y^4 + 4 \lambda y^2 \right]
\end{equation}
at one loop. For large values of the Yukawa coupling $y$ this leads to a lowering of the instability scale $\mu_I$ relative to its value in the SM.

\begin{figure}
    \centering
    \includegraphics[width=0.8\textwidth]{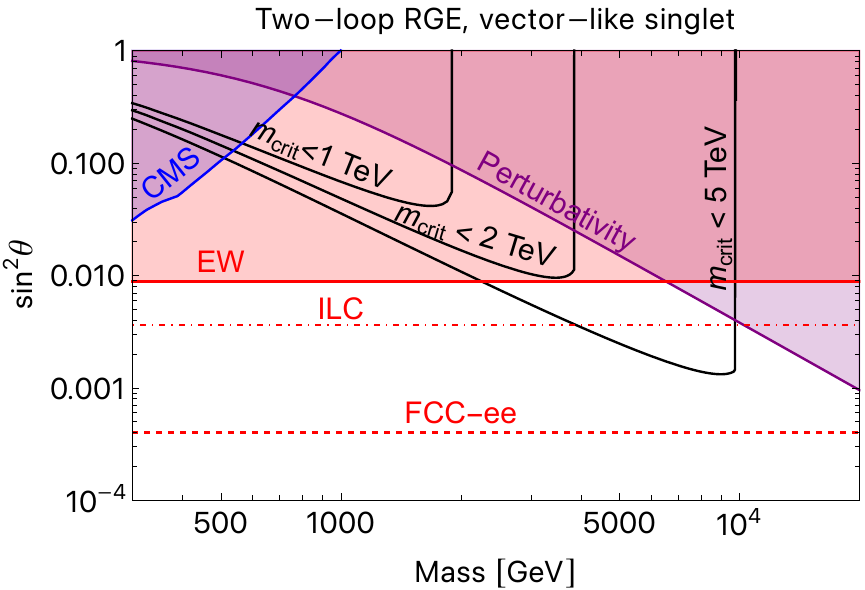}
    \caption{Higgs mass bounds and experimental constraints for the heavy neutral lepton model. The black lines show contours of constant $m_{\rm crit} = 1,2, 5$~TeV. The red solid line and shaded area correspond to experimental constraints from EW precision observables~\cite{delAguila:2008pw,deBlas:2013gla,Basso:2013jka,Deppisch:2015qwa}, and the red broken lines are projected reach from EW precision at FCC-ee (dashed)~\cite{Antusch:2015mia,Antusch:2016ejd} and ILC (dash-dotted)~\cite{Antusch:2015mia}. The blue solid line and shaded region depicts constraints from a CMS direct search~\cite{CMS:2024xdq}, and the purple solid line and shaded region correspond to constraints from requiring perturbativity of $y$.}
    \label{fig:singlet_exclusion_two_loop}
\end{figure}

In Fig.~\ref{fig:singlet_exclusion_two_loop} we show constraints on the parameter space of this model in the $(m, \sin^2 \theta)$ plane. We determine the instability scale by solving the two-loop renormalization group equations (three-loop in the gauge couplings), which we provide in appendix~\ref{sec:rge}.\footnote{We also incorporate the one-loop threshold correction to $\lambda$ and $y_t$ when matching at the scale $m$, as well as the non-logarithmic one-loop contributions to $\lambda_{\rm eff}$ in $V_{\rm eff}$. These effects are important for the large values of $y$ that we consider. For example, the one-loop threshold correction, which is parametrically $y^4/(4 \pi)^2$, can be comparable to $\lambda$. This gives rise to an $O(1)$ shift in the initial conditions for the running, which changes the instability scale (and therefore $m_{\rm crit}$) exponentially. } The black lines in Fig.~\ref{fig:singlet_exclusion_two_loop} depict contours for the maximum allowed Higgs mass parameter $m_{\rm crit}$. Patches with a heavier Higgs undergo a crunch since no metastable vacuum is formed, and we are assuming the true vacuum has a large and negative energy density. We show contours for $m_{\rm crit} = 1,2,5$~TeV. We also require that the theory be weakly coupled, i.e.\ $y < \sqrt{4\pi}$, which is depicted by the purple line in Fig.~\ref{fig:singlet_exclusion_two_loop}. We have verified that this condition is sufficient to ensure the absence of Landau poles below the instability scale.

Fig.~\ref{fig:singlet_exclusion_two_loop} presents experimental bounds on the model from EW precision tests (solid red line)~\cite{delAguila:2008pw,deBlas:2013gla,Basso:2013jka,Deppisch:2015qwa} and from a CMS search for heavy neutral leptons (blue line)~\cite{CMS:2024xdq}. EW precision observables provide the strongest constraint: one cannot make $m_{\rm crit}$ less than about $2$~TeV without running into conflict with them.\footnote{On its own, this value could suggest a residual tuning to obtain the observed Higgs mass $m \sim 100$~GeV. It is, however, crucial to recall that near-saturation of the bound on the Higgs mass would correspond to a potential with no barrier protecting the EW vacuum. In such a regime, high-energy processes in the early Universe could cause the Higgs field to leave this vacuum, leading the corresponding patches to crunch. It remains an open question how much of an additional suppression of the mass parameter is necessary to generate a large enough potential barrier to prevent this scenario.} We will shortly consider another model where $m_{\rm crit}$ can be lowered further.

Future lepton colliders can probe our parameter space through improved bounds on electroweak precision observables. We show projections for the ILC (dash-dotted red line)~\cite{Antusch:2015mia} and FCC-ee (dashed red line)~\cite{Antusch:2015mia,Antusch:2016ejd} in Fig.~\ref{fig:singlet_exclusion_two_loop}, assuming a run at the $Z$ pole. The former would probe all of the parameter space up to $m_{\rm crit} = 3$~TeV, while the latter would explore up to $m_{\rm crit} = 8.5$~TeV.

\subsection{Singlet-doublet model}
Next we consider the well-known singlet-doublet model~\cite{Mahbubani:2005pt,DEramo:2007anh,Enberg:2007rp,Cohen:2011ec,Cheung:2013dua,Abe:2014gua,Calibbi:2015nha,Freitas:2015hsa,Banerjee:2016hsk,Cai:2016sjz,LopezHonorez:2017zrd,Arcadi:2019lka,Fraser:2020dpy,Homiller:2024uxg}. This model involves one singlet, $\psi_L$, and a pair of $SU(2)$ doublets $\chi_{L,R}$ with hypercharge $1/2$. The Lagrangian is
\begin{equation}\begin{split}
    \mathcal{L}_{\psi\chi} &= i \left( \overline{\psi_L} \slashed{\partial} \psi_L + \overline{\chi_L} \slashed{D} \chi_L  + \overline{\chi_R} \slashed{D} \chi_R \right) - \left(\frac{1}{2}m_S \psi_L \psi_L + m_D \overline{\chi_R} \chi_L + {\rm ~h.c.} \right) \\
    &-y_1 \chi_L \tilde{H} \psi_L - y_2 \overline{\chi_R} H \psi_L - \lambda_i L_L^i H \psi_L - \lambda'_i \chi_L H \overline{e_R}^i + {\rm ~h.c.}
\end{split}\end{equation}
For convenience we make some further simplifying assumptions. We will assume all of the couplings to be real, so there is no CP violation in the new physics. We will also take $\lambda_i = \lambda'_i = 0$, which is technically natural due to the $\mathbb{Z}_2$ symmetry under which the new fermions are odd and SM fields are even. We define $y$ and $\theta$ such that $y_1 = y \cos \theta$ and $y_2 = y \sin \theta$, so that $y$ sets the overall scale of the couplings and $\theta$ sets their ratio.

\begin{figure}
     \centering
     \subfloat{\includegraphics[width= 0.5\textwidth]{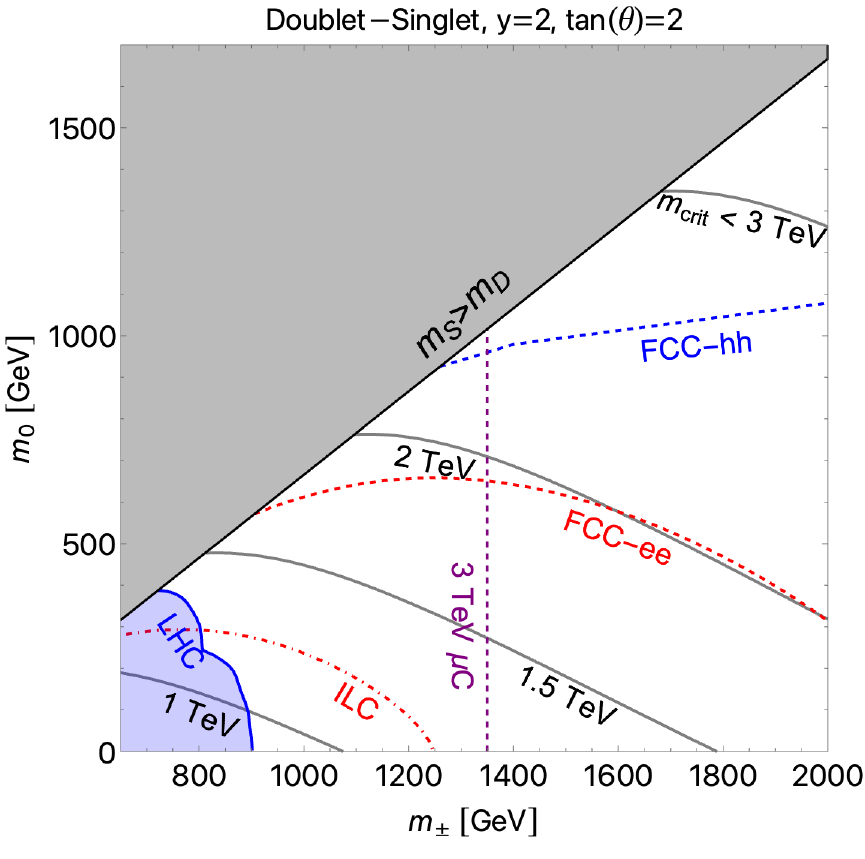}}
     \subfloat{\includegraphics[width= 0.5\textwidth]{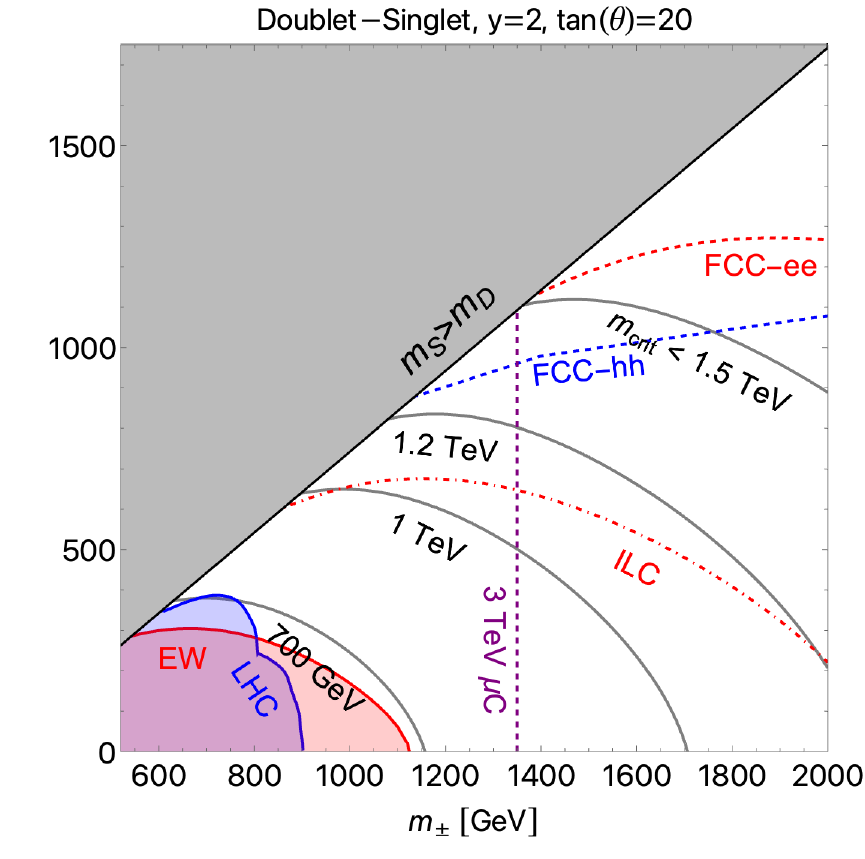}}
        \caption{Metastability bounds and experimental constraints for the doublet-singlet model taking $y=2$ and $\tan(\theta) = 2 (20)$ in the left (right) panel respectively. The labeled contours show the value of $m_{\rm crit}$ for this model. For simplicity, we consider only the region where $m_S < m_D$; the region where this condition is violated is shaded gray in both panels. The red solid line and shaded region in the right panel correspond to present-day experimental constraints from EW precision observables~\cite{Baak:2014ora}, and the red broken lines in both panels are projected reach from electroweak precision at FCC-ee (dashed)~\cite{Fan:2014vta,deBlas:2016nqo} and ILC (dash-dotted)~\cite{Fan:2014vta}. The solid blue line and shaded region shows present-day constraints from direct searches at the LHC~\cite{ATLAS:2019lff,ATLAS:2021yqv,ATLAS:2022zwa,CMS:2022sfi,ATLAS:2022hbt,CMS:2024gyw}, and the dashed blue line shows projected reach from direct searches at FCC-hh~\cite{Mangano:2017tke}. Finally, the broken purple vertical lines show projected reach of direct searches for charged particles production at a future $3 \, \text{TeV}$ muon collider~\cite{Homiller:2024uxg}.}
        \label{fig:singlet_doublet_exclusion}
\end{figure}

This model is analogous to a Higgsino-bino system, the electroweakino sector in the MSSM in the limit where the wino decouples. We provide the two-loop (three-loop for the gauge couplings) renormalization group equations that we use for this model in appendix~\ref{sec:rge}, in addition to the threshold corrections to $y_t$ and $\lambda$. Like in the heavy neutral lepton model, the addition of vector-like fermions with large Yukawa couplings lowers the instability scale of the Higgs potential.

After EW symmetry breaking, the five fermionic degrees of freedom organize into an electrically charged Dirac fermion with mass $m_\pm \equiv m_D$ and three neutral Majorana fermions. In the limit where $y_1 v, y_2 v \ll m_D, m_S$, two of these neutral fermions are pure doublets and have mass $m_D$, while the other one is a pure singlet with mass $m_S$. For generic parameters there is a nontrivial mass mixing; for explicit analytic formulae for the neutral fermion masses see, e.g., Ref.~\cite{Enberg:2007rp}. For simplicity, we focus on the regime where the doublet mass $m_D$ is larger than the singlet mass $m_S$, so that the lightest neutral particle with mass $m_0$ is singlet-dominated. In the MSSM this corresponds to assuming a bino-like LSP and a Higgsino-like NLSP. For the parameters we consider, the mass splitting $m_\pm - m_0$ is then sufficiently large that the charged fermion decays into the lightest neutral state promptly through emission of an on-shell $W$.

Fig.~\ref{fig:singlet_doublet_exclusion} depicts the parameter space of this model as a function of $m_\pm$ and $m_0$, fixing the Yukawa couplings as $y = 2$, $\tan \theta = 2, 20$. We include bounds from a number of direct searches for electroweakinos at the LHC~\cite{ATLAS:2019lff,ATLAS:2021yqv,ATLAS:2022zwa,CMS:2022sfi,ATLAS:2022hbt,CMS:2024gyw}, which are compiled into the blue line in Fig.~\ref{fig:singlet_doublet_exclusion}. EW precision bounds~\cite{Baak:2014ora} are indicated by the solid red line, where we compute the $S$ and $T$ parameters following Ref.~\cite{DEramo:2007anh}.\footnote{We find that the constraints resulting from contributions to $S$ are subleading to existing bounds in all cases we consider. However, in the $\tan \theta = 1$ limit, these contributions could be relevant for EW precision searches at future colliders.}

This model is harder to probe experimentally than the heavy neutral lepton model in the previous section. Consequently we obtain a stronger bound on the Higgs mass while still being consistent with experimental constraints. Fig.~\ref{fig:singlet_doublet_exclusion} illustrates that the upper bound on the Higgs mass can be as low as $m_{\rm crit} \sim 700$~GeV.

Future lepton colliders can effectively probe this model, both through EW precision observables and through direct searches for charged particle pair production. In Fig.~\ref{fig:singlet_doublet_exclusion} we again show projections for EW precision bounds at the ILC (dash-dotted red line)~\cite{Fan:2014vta} and the FCC-ee (dashed red line)~\cite{Fan:2014vta,deBlas:2016nqo}. For direct searches we show a projection for a $\sqrt{s} = 3$~TeV muon collider (dashed purple line) in Fig.~\ref{fig:singlet_doublet_exclusion}. We assume that charged particles can be discovered for masses up to $0.45\sqrt{s}$, following the rule of thumb in Ref.~\cite{AlAli:2021let}. We also include a projection for FCC-hh (dashed blue line)~\cite{Mangano:2017tke}.

The Higgs decay widths to EW gauge bosons, $h \rightarrow WW^*$ and $h \rightarrow Z Z^*$, deviate from the SM prediction. The present bounds on Higgs signal strengths are not strong enough to provide an important constraint in Fig.~\ref{fig:singlet_doublet_exclusion}. The contribution of the new fermions to the $WW$ signal strength scales as $\delta \mu_{hWW} \sim y^4 v^2 / 16\pi^2 m^2$, where $m$ is the mass scale of the fermions~\cite{DAgnolo:2023rnh}, while the signal strength is measured to about $9\%$~precision at the LHC~\cite{CMS:2022dwd}. For TeV-scale masses we estimate that this constraint is only important for $y \gtrsim 4$, which is in rough agreement with the more careful computations in~\cite{Homiller:2024uxg}. The FCC-ee or a muon collider could potentially probe $\mu_{hWW}$ to $\sim 0.4\%$ precision, which could plausibly be sensitive to the parameter space in Fig.~\ref{fig:singlet_doublet_exclusion}~\cite{Homiller:2024uxg,Bernardi:2022hny,Forslund:2022xjq}. We leave a detailed study of this to future work.

\section{Vacuum decay}\label{sec:cosmology}
Our mechanism implies the metastability of the EW vacuum, while obtaining a Higgs mass bound close to the observed value requires unrelated new physics capable of drastically destabilizing it. Ensuring that our patch is sufficiently long-lived to be consistent with the observed age of the Universe therefore requires additional physics capable of (partially) stabilizing the vacuum. In the following, we provide a rough upper bound on the scale $\Lambda_{\rm stab}$ where this new physics can be expected to become relevant. For a detailed analysis of this stabilizing effect, see Ref.~\cite{Enguita:2025ybx}. These authors perform a complete numerical analysis for a concrete model, showing that the scale of stabilizing physics can indeed be significantly lower than (and hence in compliance with) the analytical bound we obtain this section. This improvement is indeed sufficiently strong to place this new physics in range of FCC-hh.

In appendix~\ref{sec:reheatingdecay} we review important aspects of the computation of the decay rate. If the scale of new physics is high enough for its stabilizing effects to be negligible, these considerations give rise to a lower bound on the quartic coupling $\lambda$,
\begin{equation}\label{eq:quarticbound}
    \lambda \gtrsim -0.06 \quad {\rm (no~stabilization)} , 
\end{equation} 
in agreement with~\cite{Isidori:2001bm}.

An important subtlety of this result is its RG scale-dependence. In Ref.~\cite{Andreassen:2017rzq}, it was shown that in the limit of a negligible stabilizing effect from new physics, the correct scale choice is $\mu=\mu_*$, the scale at which the quartic coupling reaches its minimum, $\beta_\lambda (\mu_*)=0$~\cite{Andreassen:2017rzq}. If the new physics stabilizes the vacuum only through its effects on the running of $\lambda$ these results still apply. This implies that this physics needs to become active at a scale $\Lambda_{\rm stab}$ smaller than the scale at which $\lambda$ saturates Eq.~\eqref{eq:quarticbound}. In this case, the scale $\Lambda_{\rm stab}$ can a priori be different from the scale $\Lambda$ associated with the natural vacua. 

If, on the other hand, the stabilization is based on a modification of the tree-level potential, these additional terms will also influence the choice of scale in a nontrivial way. As we argue in appendix~\ref{sec:reheatingdecay}, the perturbative scheme laid out in Refs.~\cite{Khoury:2021zao,Steingasser:2022yqx,Chauhan:2023pur} can be used to constrain $\Lambda_{\rm stab}$ indirectly. To leading order, the corrections to the potential lower the relevant scale down to the so-called \textit{instanton scale} $\mu_S$. For a fast-running $\lambda$, this scale lies roughly one order of magnitude below $\Lambda_{\rm stab}$. In this regime, this procedure neglects important contributions to $\Lambda_{\text{stab}}$ from the full potential. While a more reliable analysis could be obtained by specifying a concrete UV completion, we will here restrict ourselves to obtaining a rough estimate relying on this perturbative scheme.

To illustrate the ramifications of Eq.~\eqref{eq:quarticbound} for our mechanism, we can consider the one-loop estimate for the running quartic coupling. We approximate the $\beta$ function as a constant near the instability scale, so that the running coupling is $\lambda(h) = b \log \frac{h}{\mu_I}$. New physics should become important before the running coupling reaches $-0.06$ to stabilize the vacuum, leading to 
\begin{equation}\label{eq:lifetimebound}
    \mu_S \lesssim \mu_I e^{0.06/\abs{b}}.
\end{equation}
As a benchmark point, let us consider the heavy neutral lepton model with $M = 3$~TeV and the Higgs mass bound $m_{\rm crit} = 5$~TeV. From Fig.~\ref{fig:singlet_exclusion_two_loop} one can see that this requires a mixing angle $\sin^2\theta \approx 6 \times 10^{-3}$, which is consistent with experimental constraints. From Eq.~\eqref{eq:mixingangle} we find that the Yukawa coupling for this benchmark is $y \approx 1.3$. The $\beta$ function can then be approximated (see Eq.~\eqref{eq:onelooprg}) as $b \approx  -0.04$. The instability scale can be determined using Eq.~\eqref{eq:mcrit}, leading to $\mu_I \approx 30$~TeV. Finally, applying Eq.~\eqref{eq:lifetimebound} gives a bound $\mu_S \lesssim 140$~TeV, and hence $\Lambda_{\rm stab}\lesssim 10^3$~TeV. Some new physics needs to become relevant below this scale to stabilize the vacuum. 

The stabilizing physics could either involve a tree-level modification of the potential or modify the running of the quartic. In the former case, one expects $\Lambda_{\rm stab}$ to be of the same order as the scale of the natural vacuum $\Lambda$~\cite{Detering:2024vxs,Enguita:2025ybx,Detering:2025nmh}. An example of this is the Majoron model of neutrino masses~\cite{Chikashige:1980ui}. For the case of a stabilization through a modified running, one can have $\Lambda \gg \Lambda_{\rm stab}$. For example, the theory could approach a near-conformal fixed point in the UV such that the quartic approaches a constant greater than $-0.06$~\cite{Giudice:2021viw}.

\section{Conclusions}\label{sec:conclusions}
The apparent fine-tunings of the SM, manifesting in the cosmological constant problem, the metastability of the electroweak vacuum, and the (gauge) hierarchy problem, pose one of the greatest challenges of modern fundamental physics. In this article, we presented a novel perspective on these issues by combining the ideas of dynamical vacuum selection and metastability bounds on the Higgs mass. 

In accordance with the idea of vacuum selection, we assumed a landscape where the parameters of the Higgs potential in each patch are set by some scanning sector. Regarding this sector, we assumed that it generically gives rise to an effective Higgs potential with at least one minimum, in which the Higgs takes a natural VEV with negative vacuum energy density. We have then shown that if the Higgs mass parameter is small enough, running effects can lead to the formation of an additional, metastable vacuum with significantly higher energy density. Remarkably, this simple observation offers an explanation for both the metastability of the EW vacuum as well as the hierarchy problem. Furthermore, its ability to explain the observed Higgs mass hinges on the existence of new physics to lower the instability scale. Conversely, if such new physics is observed, this discovery could serve as circumstantial evidence that a scanning mechanism is indeed realized in nature.
Our results can also be understood as a small Higgs mass allowing for the formation of a feasible vacuum within a landscape of generically infeasible vacua. While in our work we realized this through the crunching dynamics, it would be interesting to explore other possibilities.

Adding to previous works that have studied ways to lower the instability scale~\cite{Giudice:2021viw,Khoury:2021zao,Benevedes:2024tdq,Detering:2024vxs}, we provided a detailed phenomenological assessment of two new realizations of this idea: heavy neutral leptons and the singlet-doublet model. These models exhibit distinct phenomenology, different from symmetry-based solutions and other vacuum selection mechanisms. They predict vector-like fermions which can be observed at future lepton colliders such as the FCC-ee or a muon collider. 

Lastly, since our argument relies on a dynamical selection mechanism rather than a statistical one, our conclusion of an unnaturally small Higgs mass does not directly depend on a measure. Moving forward, the conditions necessary for our analysis can serve as guiding principles for the development of concrete, testable vacuum selection models. As we generically expect UV completions to involve new bosons to partially stabilize the vacuum, it would be particularly interesting to investigate their signatures in collider physics and cosmology.

\acknowledgments
We are grateful to Raffaele Tito D'Agnolo and Jesse Thaler for providing helpful comments on our manuscript. We thank the organizers of DPF-PHENO 2024, where this work was initiated. AI is supported by a Mafalda and Reinhard Oehme Postdoctoral Research Fellowship from the Enrico Fermi Institute at the University of Chicago. Portions of this work were conducted in MIT's Center for Theoretical Physics and partially supported by the U.S. Department of Energy under Contract No.~DE-SC0012567. This project was also supported in part by the Black Hole Initiative at Harvard University, with support from the Gordon and Betty Moore Foundation and the John Templeton Foundation. The opinions expressed in this publication are those of the author(s) and do not necessarily reflect the views of these Foundations.

\appendix

\section{Physical Higgs mass bound}\label{sec:stepfunction}
Here we extend the analysis in section~\ref{sec:OLE} to establish a bound on the physical Higgs mass $M_H^2$. We will find that it is bounded from above by $e^{-1} m_{\rm crit}^2$ for a constant $\beta$ function, where $m_{\rm crit}^2$ is given in Eq.~\eqref{eq:mcrit}.

With a constant $\beta$ function we showed the Higgs mass \textit{parameter} must be smaller than $m_{\rm crit}^2$ for the metastable vacuum to exist. This bound can also be stated in terms of the Higgs VEV $v$.
The Higgs VEV arising from a critical mass parameter is given by
\begin{equation}\label{eq:vcrit}
    v_{\rm crit} = e^{-3/4} \mu_I ,
\end{equation}
and the metastable vacuum exists only if $v < v_{\rm crit}$. In this case, the VEV is given by an inflection point rather than a local minimum.

By expanding the potential around the VEV we identify the physical mass of the Higgs as
\begin{equation}
    M_H^2 = -2 \abs{\beta_\lambda} v^2 \log \frac{v}{v_{\rm crit}} = m_{\rm crit}^2 \frac{v^2}{v_{\rm crit}^2} \log \frac{v_{\rm crit}^2}{v^2} .
\end{equation}
This vanishes both at $v = 0$ and at the critical value $v = v_{\rm crit}$, as expected. The Higgs mass is maximized at $v_* = e^{-1/2} v_{\rm crit} = e^{-5/4} \mu_I$, leading to a bound
\begin{equation}
    M_H^2 \leq e^{-1} m_{\rm crit}^2  \quad ({\rm constant~} \beta_\lambda).
\end{equation}

\section{Upper bound on the stabilization scale}\label{sec:reheatingdecay}

\subsection{Zero-temperature estimate}
In field theory, the decay of a false vacuum proceeds through the formation of a bubble of true vacuum. The rate of this process per unit volume is of the form
\begin{align}
    \frac{\Gamma}{V}\sim \mu_S^4 e^{-S_E [\phi_I]},
\end{align}
where $S_E$ denotes the Euclidean action of the instanton $\phi_I$. This configuration is defined as a solution to the equations of motion in Euclidean time $\tau$ subject to the appropriate boundary conditions. The typical scale of this configuration corresponds to $\mu_S$. The ongoing existence of the EW vacuum implies that the probability that a vacuum bubble has been nucleated within our past lightcone $\mathcal{P}$ should be smaller then $1$,
\begin{equation}
	 \int_{\mathcal{P}} \text{d}^4 x \ \frac{\Gamma}{V}\ll 1.
    \label{eq:total-probability}
\end{equation}
Current cosmological data suggest that the Universe is roughly 14 billion years old, and that it has been matter-dominated for a vast majority of this time. This makes it straightforward to obtain the volume of $\mathcal{P}$ as~\cite{Buttazzo:2013uya}
\begin{equation}
    V_{\mathcal{P}}=\frac{0.15}{H_0^4}=2.2 \times 10^{163}(\text{GeV})^{-4},
    \label{eq:VolumeLightcone}
\end{equation}
using the observed value of the cosmological constant, $H_0 \approx 67.4 \frac{\rm km}{\rm s \cdot Mpc}$. This translates to a lower bound on the Euclidean action of the instanton,
\begin{equation}\label{eq:SEbound}
    S_E>410 + 4 \ln \left( \frac{\mu_S}{\text{TeV}} \right).
\end{equation}
To understand how this relates to the scenario discussed in this work, consider the SM extended by vector-like fermions. The approximate scale-invariance of the scalar sector leads to the existence of not one, but an infinite number of instantons forming a one-parameter family $\phi_{I,R}$, where $R$ controls the typical field values as well as the instanton's size. For each of these instantons, the Euclidean action is known to be given by the simple expression~\cite{Linde:1981zj,Isidori:2001bm,Buttazzo:2013uya}
\begin{equation}\label{eq:thickwallbounceaction}
    S_E = \frac{8\pi^2}{3 \abs{\lambda}} . 
\end{equation}
The scale invariance of the potential is broken by the RG running of $\lambda$. In Ref.~\cite{Andreassen:2017rzq}, it was shown that the effect of the running is to pick out a scale in Eq.~\eqref{eq:thickwallbounceaction} at which $\lambda$ should be evaluated, $\lambda=\lambda (\mu_*)$.
Here, $\mu_*$ is the scale where $\lambda$ reaches its minimum. The vector-like fermions destabilize the vacuum by driving $\lambda$ more negative. Eqs.~\eqref{eq:SEbound} and~\eqref{eq:thickwallbounceaction} then imply that requiring the lifetime to exceed the age of the Universe sets a lower limit on $\lambda$. Taking $\mu_S \sim \mathcal{O}$(TeV), this bound is given by $\lambda \gtrsim -0.06$~\cite{Isidori:2001bm}. 

In the context of our mechanism, this condition would generically be violated unless additional new physics partially stabilizes the vacuum --- either by changing the running of $\lambda$, or by modifying the tree-level potential. For the first case, the scale of the stabilizing physics $\Lambda_{\rm stab}$ is bounded from above by the scale where $\lambda$ becomes smaller than $-0.06$. For the latter case, it is well-understood that the interplay between the RG running of $\lambda$ and higher-dimensional operators lead to a nontrivial change of the relevant scale~\cite{Khoury:2021zao}. In particular, for the relatively fast running of $\lambda$ relevant to our discussion, this effect renders the tunneling rate highly dependent on the details of the full potential of the UV completion.

For the sake of obtaining a rough upper bound on $\Lambda_{\rm stab}$, however, we can rely on the perturbative scheme laid out in Refs.~\cite{Khoury:2021zao,Steingasser:2022yqx,Chauhan:2023pur}. In this picture, the tunneling rate is still controlled by the quartic coupling $\lambda$, but $\lambda$ should be evaluated at the so-called \textit{instanton scale}~$\mu_S$.
In our regime of interest, $\mu_S$ is expected to lie roughly one order of magnitude below $\Lambda_{\rm stab}$. Further neglecting the (stabilizing) effects of the higher-dimensional operators on the Euclidean action, we estimate that the instanton scale $\mu_S$ needs to satisfy $\lambda (\mu_S) \gtrsim -0.06$, while $\Lambda_{\rm stab}\sim \mathcal{O}(10) \cdot \mu_S$.

\subsection{Finite-temperature estimate}
In the early universe, when the temperature is above the EW scale, we must consider the effective Higgs potential at finite temperature to understand the fate of the metastable vacuum. At temperatures $T \gg M_H$, the effective Higgs mass parameter derived from the thermal effective potential is $m_{\rm eff}^2 = a T^2$, where the coefficient $a$ depends on the particle content. In the SM $a \approx 0.4$ (with a mild $T$-dependence from the running of gauge couplings), while the addition of vector-like fermions in our mechanism generally increases $a$~\cite{Katz:2014bha}.

At high temperatures $T\gg  \mu_I$, the decay of the vacuum is dominated by the essentially classical nucleation of $O(3)$-symmetric bubbles~\cite{Langer:1967ax,Langer:1969bc,Bochkarev:1992rh,Boyanovsky:1993ki}, while finite-temperature \textit{quantum} tunneling is dominant only for a relatively small range of temperatures~\cite{Steingasser:2023gde,Steingasser:2024ikl}. The $O(3)$ bubble nucleation rate is then given by
\begin{equation}
    \frac{\Gamma}{V} \sim T^4 e^{-E_b/T},
\end{equation}
where $E_b$ is the energy of the bubble. The decay proceeds efficiently if $\Gamma/V \gtrsim H^4 \sim T^8/M_{\rm Pl}^4$. This implies a lower bound on the energy of the bubble in order to avoid a phase transition,
\begin{equation}
    \frac{E_b}{T} \gtrsim 4 \log \frac{M_{\rm Pl}}{T} .
\end{equation}
Again restricting ourselves to leading-order estimates, we can now focus on operators in the potential of up to dimension four. For this scenario, it was shown in Ref.~\cite{Linde:1981zj,Salvio:2016mvj} that the energy of the bubble is given by
\begin{equation}
    \frac{E_b}{T} \approx 19 \frac{\mu_{\rm eff}}{\abs{\lambda(T)} T} = 19 \frac{\sqrt{a}}{\abs{\lambda(T)}} ,
\end{equation}
where the factor of $19$ arises from a numerical calculation of the bubble profile with a constant $\lambda$. Imposing $S_3 / T > 150$ and taking $a = 0.4$ yields a bound on the quartic of $\lambda \gtrsim -0.08$. This is weaker than the bound we derived from considering false vacuum decay at low temperature, $\lambda \gtrsim -0.06$.

\section{Two-loop RGE}\label{sec:rge}
In this appendix, we provide the renormalization group equations (RGE) for the heavy neutral lepton model and for the singlet-doublet model considered in section~\ref{sec:models}. We compute the $\beta$ functions using the \texttt{RGBeta} package~\cite{Thomsen:2021ncy}, which calculates $\beta$ functions to two-loop order for scalar and Yukawa couplings and to three-loop order for gauge couplings. We also compute the one-loop threshold corrections to $\lambda$ and $y_t$ for each of these models using the \texttt{Matchete} package~\cite{Fuentes-Martin:2022jrf}, matching at the scale $M$ for the the heavy neutral lepton model and at the scale $m_D$ for the singlet-doublet model.

We tabulate the $\beta$ functions and threshold corrections for the heavy neutral lepton model in subsection~\ref{subsec:hnlrge} and for the singlet-doublet model in subsection~\ref{subsec:sdrge}. For simplicity we introduce the following auxiliary parameters,
\begin{align}
    Y_2 \equiv &  y_t^2 +  y_b^2 - \frac{1}{3} y_\tau^2\\
    Y_4 \equiv &  y_t^4 +  y_b^4 - \frac{1}{3} y_\tau^4\\
    Y_6 \equiv &  y_t^6 +  y_b^6 - \frac{1}{3} y_\tau^6.
\end{align}
Below the scale of new physics, we use the following equations for the SM RGE. The running of Higgs' quartic coupling is controlled by
\allowdisplaybreaks
\begin{align}
    \frac{\text{d}\lambda}{\text{d}\ln \mu} =\frac{1}{(4 \pi)^2}\bigg[& 24 \lambda ^2 - 6 Y_2 +\frac{9 g^4}{8} +\frac{3 \left(g'\right)^4}{8}+\frac{3}{4} g^2 \left(g'\right)^2 +\lambda \left( 12 Y_2 -9 g^2 -3 \left(g'\right)^2 \right)\bigg]+\nonumber\\
    +\frac{1}{(4 \pi)^4}\bigg[&30 Y_6  -6 (y_b^2 y_t^4 + y_b^4 y_t^2 ) -32 g_3^2 (y_t^4 + y_b^4 ) + (g^\prime)^2 \left(-\frac{8}{3} y_t^4 + \frac{4}{3} y_b^4 -4 y_{\tau }^4\right) \nonumber \\
    & - \frac{9}{4} g^4  Y_2 + \frac{1}{4} (g^\prime)^4 \left( -19 y_t^2 + 5 y_b^2-25y_\tau^2 \right) \nonumber\\
    &  +  \frac{1}{2} g^2 (g^\prime)^2 \left( 21 y_t^2 + 9 y_b^2 + 11 y_{\tau }^2\right) +\frac{305 g^6}{16} -\frac{379 \left(g'\right)^6}{48} -\frac{289}{48} g^4 \left(g'\right)^2 \nonumber\\
    &  -\frac{559}{48} g^2 \left(g'\right)^4 +\lambda y_t^2  \left(\frac{45}{2} g^2 +\frac{85}{6} \left(g'\right)^2  +80 g_3^2  \right) \nonumber\\
    &+\lambda y_b^2  \left(\frac{45}{2} g^2  +\frac{25}{6} \left(g'\right)^2 +80 g_3^2 \right)+ \lambda  y_{\tau }^2 \left(\frac{25}{2} \left(g'\right)^2 \right) \nonumber\\
    &+\lambda \bigg(  -3 Y_4 -42 y_b^2 y_t^2 +\frac{15}{2} g^2 y_{\tau }^2+\frac{629}{24}  \left(g'\right)^4 -\frac{73 g^4}{8} +\frac{39}{4} g^2 \left(g'\right)^2  \bigg) \nonumber\\ 
   & +\lambda^2 \left(-144 Y_2 +36  \left(g'\right)^2 + 108 g^2\right) -312 \lambda^3 \bigg].
\end{align}
For the Yukawa couplings, we have
\begin{align}
    \frac{\text{d}y_t}{\text{d}\ln \mu} = \frac{y_t}{(4 \pi )^2}\bigg[& \frac{9 y_t^2}{2} + \frac{3}{2} y_b^2 + y_{\tau }^2 -\frac{9 g^2 }{4}-\frac{17}{12} \left(g'\right)^2 -8 g_3^2 \bigg] \nonumber\\ 
    +\frac{y_t}{(4 \pi)^4}\bigg[&6 \lambda ^2 -12 \lambda  y_t^2 -12 y_t^4 -\frac{1}{4} y_b^4 -\frac{9}{4}  y_{\tau }^4 -\frac{11}{4} y_t^2 y_b^2  -\frac{9}{4} y_t^2 y_{\tau }^2 +\frac{5}{4} y_b^2 y_{\tau }^2 \nonumber\\
    &+ g^2 \left( \frac{225}{16} y_t^2 +\frac{99}{16} y_b^2 +\frac{15}{8} y_{\tau }^2\right) + (g^\prime)^2 \left(\frac{131}{16} y_t^2 +\frac{7}{48} y_b^2  +\frac{25}{8} y_{\tau }^2 \right)\nonumber \\ 
    & +g_3^2 (36  y_t^2 + 4 y_b^2 ) -\frac{23 g^4 }{4}+\frac{1187}{216} \left(g'\right)^4 -108 g_3^4  -\frac{3}{4} g^2 \left(g'\right)^2\nonumber \\
    & +9 g^2 g_3^2 +\frac{19}{9} g_3^2  \left(g'\right)^2\bigg], \\
    \frac{\text{d}y_b}{\text{d}\ln \mu} = \frac{y_b}{(4 \pi)^2}\bigg[ & \frac{3}{2} y_t^2+\frac{9 y_b^2}{2}+ y_{\tau }^2-\frac{9 g^2 }{4}-\frac{5}{12} \left(g'\right)^2-8 g_3^2 \bigg]\nonumber\\
    +\frac{y_b}{(4 \pi)^4}\bigg[& 6 \lambda^2 -12 \lambda  y_b^2 -12 y_b^4 -\frac{1}{4} y_t^4 -\frac{9}{4}  y_{\tau }^4 -\frac{11}{4} y_t^2 y_b^2 +\frac{5}{4}  y_t^2 y_{\tau }^2 -\frac{9}{4} y_b^2 y_{\tau }^2 \nonumber \\ 
    &+ g^2 \left(\frac{99}{16}  y_t^2 +\frac{225}{16} y_b^2 +\frac{15}{8} y_{\tau }^2\right) + (g^\prime)^2 \left( \frac{91}{48}  y_t^2 +\frac{79}{16} y_b^2  +\frac{25}{8} y_{\tau }^2 \right) \nonumber \\
    &+ g_3^2 (4 y_t^2 +36 y_b^2)-\frac{23 g^4}{4}-\frac{127}{216} \left(g'\right)^4 -108 g_3^4 -\frac{9}{4} g^2 \left(g'\right)^2 \nonumber \\
    &+9 g^2 g_3^2+\frac{31}{9} g_3^2 \left(g'\right)^2\bigg] , \ \ \text{and} \\
\frac{\text{d}y_\tau}{\text{d}\ln \mu} = \frac{y_\tau}{(4 \pi)^2}\bigg[&3 y_t^2 + 3 y_b^2 + \frac{5 y_{\tau }^2}{2}  -\frac{9 g^2 }{4}-\frac{15}{4} \left(g'\right)^2 \bigg] \nonumber\\ 
    +\frac{y_\tau}{(4 \pi)^4}\bigg[& 6 \lambda^2  -12 \lambda  y_{\tau }^3 -\frac{27}{4} ( y_t^4 + y_b^4) -3 y_{\tau }^4 +\frac{3}{2} y_b^2 y_t^2  -\frac{27}{4} ( y_t^2 + y_b^2  )y_\tau^2 \nonumber \\
    & g^2 \left( \frac{45}{8} (y_t^2 + y_b^2 ) +\frac{165}{16} y_{\tau }^2 \right) + \left(g'\right)^2 \left( \frac{85}{24} y_t^2 +\frac{25}{24} y_b^2  +\frac{179}{16} y_{\tau }^2\right)  \nonumber\\
    &+20 g_3^2 (y_t^2 + y_b^2 ) -\frac{23 g^4 }{4}  +\frac{457}{24} \left(g'\right)^4 +\frac{9}{4} g^2 \left(g'\right)^2  \bigg].
\end{align}
Finally, the beta functions of the gauge couplings are
\begin{align}
\nonumber
\frac{\text{d}g^\prime}{\text{d}\ln \mu}=& \frac{{g^{\prime}}^3}{(4 \pi)^2} \frac{41}{6} + \frac{{g^{\prime}}^3}{(4 \pi)^4}\bigg[ \frac{199}{18}{g^{\prime}}^2 + \frac{9}{2}g^2 + \frac{44}{3} g_s^2 - \frac{17}{6} y_{\rm t}^2 - \frac{1}{2} |Y|^2 \bigg]  \\ 
\nonumber
&+ \frac{{g^{\prime}}^3}{(4 \pi)^6}\bigg[ y_{\rm t}^2 \bigg( \frac{315}{16} y_{\rm t}^2 -\frac{29}{5} g_s^2 - \frac{785}{32} g^2 - \frac{2827}{288} {g^\prime}^2 \bigg) + \lambda \bigg(-3 \lambda + \frac{3}{2} g^2 + \frac{3}{2} {g^\prime }^2 \bigg)  \\ 
&  + 99 g_s^4 + \frac{1315}{64} g^4 - \frac{388613}{5184} {g^\prime}^4 - \frac{25}{9} g_s^2 g^2 - \frac{137}{27} g_s^2 {g^\prime}^2 +\frac{205}{96} g^2 {g^\prime}^2 \bigg] . \\
\frac{\text{d}g}{\text{d}\ln \mu} =& - \frac{g^3}{(4 \pi)^2}\frac{19}{6}  + \frac{g^3}{(4 \pi)^4}\bigg[ \frac{3}{2}{g^{\prime}}^2 + \frac{35}{6}g^2 + 12 g_s^2 - \frac{3}{2} y_{\rm t}^2 - \frac{1}{2} |Y|^2 \bigg] \nonumber \\ 
&+ \frac{g^3}{(4 \pi)^6}\bigg[ y_{\rm t}^2 \left( \frac{147}{16} y_{\rm t}^2 - 7 g_s^2 - \frac{729}{32} g^2 - \frac{593}{96} {g^\prime}^2 \right) + \lambda \left( - 3 \lambda + \frac{3}{2} g^2 + \frac{1}{2} {g^\prime}^2 \right) \nonumber \\ 
&  + 81 g_s^4 + \frac{324953}{1728} g^4 - \frac{5597}{576} {g^\prime}^4 + 39 g_s^2 g^2 - \frac{1}{3} g_s^2 {g^\prime}^2 + \frac{291}{32} g^2 {g^\prime}^2 \bigg],  \ \ \text{and}\\
\nonumber
\frac{\text{d}g_3}{\text{d}\ln \mu} =&- \frac{g_3^3}{(4 \pi)^2} 7  + \frac{g_3^3}{(4 \pi)^4}\bigg[ \frac{11}{6}g_3^2 + \frac{9}{2}g^2 - 26 g_3^2 - 2 y_{\rm t}^2 \bigg] \\ 
\nonumber
&+  \frac{g_3^3}{(4 \pi)^6} \bigg[y_{\rm t}^2 \left(15 y_{\rm t}^2 - 40 g_3^2 - 93/8 g^2 - 101/24 {g^\prime}^2\right) \\ 
&  + \frac{65}{2} g_3^4 + \frac{109}{8} g^4 - \frac{2615}{216} {g^\prime}^4 +  21 g_3^2 g^2 + \frac{77}{9} g_3^2 {g^\prime}^2 - \frac{1}{8} g^2 {g^\prime}^2 \bigg].
\end{align}
For our renormalization group evolution, we use the initial conditions at the top mass provided in Ref.~\cite{Buttazzo:2013uya}.

\subsection{Heavy neutral lepton RGE}\label{subsec:hnlrge}
Here, we report the RGE for the heavy neutral lepton model. Since the new fermion is a gauge singlet, the RGE reduce to the SM expressions tabulated above in the $y \to 0$ limit, so for the sake of brevity, we explicitly list only the contributions arising from the new physics:
\begin{align}
    \frac{\text{d}\lambda}{\text{d} \ln \mu} \supset 
    \frac{1}{(4 \pi)^2} \bigg[& 4 y^2 \lambda - 2 y^4 \bigg]  \\
    +\frac{y^2}{(4 \pi)^4} \bigg[&10 y^4 -\frac{3}{4} g^4+\frac{15}{2} g^2 \lambda - \frac{1}{4}\left(g'\right)^4+ \frac{7}{2} \left(g'\right)^2 g^2+\frac{5}{2} \lambda \left(g'\right)^2- 48 \lambda ^2-\lambda  y^2 \bigg], \nonumber\\
    \frac{\text{d}y_t}{\text{d} \ln \mu} \supset \frac{y_t}{(4 \pi)^2} \big[& y^2 \big]  + \frac{y_t}{(4 \pi)^4} \bigg[ \frac{5}{4} y^2 y_b^2 +\frac{15}{8} g^2 y^2 +\frac{5}{8} y^2 \left(g'\right)^2 -\frac{9}{4}  y^4 -\frac{9}{4} y^2 y_t^2 \bigg] ,\\
    \frac{\text{d}y_b}{\text{d} \ln \mu} \supset \frac{y_b}{(4 \pi)^2} \big[& y^2 \big] + \frac{y_b}{(4 \pi)^4} \bigg[ \frac{15}{8} g^2 y^2 +\frac{5}{8} y^2  \left(g'\right)^2+\frac{5}{4} y^2  y_t^2-\frac{9}{4}  y^4 -\frac{9}{4} y^2 y_b^2 \bigg], \\
    \frac{\text{d}y_\tau}{\text{d} \ln \mu} \supset \frac{y_\tau}{(4 \pi)^2} \big[ & y^2 \big] + \frac{y_\tau}{(4 \pi)^4} \bigg[\frac{15}{8} g^2 y^2 +\frac{5}{8} y^2 \left(g'\right)^2 -\frac{9}{4} y^4 -\frac{9}{4} y^2 y_{\tau}^2 \bigg], \\
    \frac{\text{d}g^\prime}{\text{d} \ln \mu} \supset \frac{1}{(4 \pi)^4} \bigg[& -y^2 \left(g'\right)^4 \bigg] +\frac{1}{(4 \pi)^6} \bigg[ \frac{31}{6} y^2 y_b^2 \left(g'\right)^4+\frac{73}{6} y^2 \left(g'\right)^4 y_t^2+\frac{45}{8} y^4 \left(g'\right)^4 \nonumber\\ 
    &+\frac{19}{2} y^2 \left(g'\right)^4 y_{\tau }^2-\frac{65}{16}  y^2 \left(g'\right)^6-\frac{195}{16} g^2 y^2 \left(g'\right)^4\bigg], \\
    \frac{\text{d}g}{\text{d} \ln \mu} \supset \frac{1}{(4 \pi)^4} \bigg[& -g^4 y^2\bigg]+ \frac{1}{(4 \pi)^6} \bigg[ \frac{15}{2} g^4 y^2 y_b^2-\frac{243}{16} g^6 y^2+\frac{15}{2} g^4 y^2 y_t^2+\frac{29 g^4 y^4}{8}\nonumber\\ 
    &+\frac{5}{2} g^4 y^2 y_{\tau }^2-\frac{65}{16} g^4 y^2 \left(g'\right)^2\bigg], \\ 
    \frac{\text{d}g_3}{\text{d} \ln \mu} \supset \frac{1}{(4 \pi)^6}  \bigg[&  7 y^2 g_3^4 (y_t^2+y_b^2 ) \bigg].
\end{align}
For the additional Yukawa coupling, we find
\begin{align}
    \frac{\text{d}y}{\text{d} \ln \mu} = \frac{y}{(4 \pi)^2} \bigg[& 3  Y_2  +\frac{5 y^2}{2} -\frac{9 g^2 }{4}-\frac{3}{4}  \left(g'\right)^2 \bigg] \nonumber \\ 
+\frac{y}{(4 \pi)^4} \bigg[& \frac{45}{8} g^2  y_b^2+\frac{25}{24}  y_b^2 \left(g'\right)^2+20 g_3^2  y_b^2+\frac{3}{2}  y_b^2 y_t^2-\frac{27}{4} y^2 y_b^2-\frac{27  y_b^4}{4}-\frac{23 g^4 }{4} \nonumber\\ 
&+\frac{45}{8} g^2 y_t^2+\frac{165 g^2 y^2}{16}+\frac{15}{8} g^2  y_{\tau }^2+\frac{85}{24}  \left(g'\right)^2 y_t^2+\frac{103}{16} y^2 \left(g'\right)^2 \nonumber \\ 
&+\frac{25}{8}  \left(g'\right)^2 y_{\tau }^2+\frac{35}{24}  \left(g'\right)^4-\frac{9}{4} g^2  \left(g'\right)^2+20 \text{g3}^2  y_t^2-\frac{27}{4} y^2 y_t^2-\frac{27  y_t^4}{4} \nonumber\\ 
&-3 y^4-12 \lambda  y^2-\frac{9}{4} y^2 y_{\tau }^2+6 \lambda ^2 -\frac{9  y_{\tau }^4}{4} \bigg] .
\end{align}
We furthermore find the following threshold corrections matching at the scale $M$:
\begin{align}
    \lambda_{\rm UV} - \lambda_{\rm IR}&=\frac{y^2 \lambda_{\rm IR}}{16 \pi^2}, \\ 
    y_{t,\rm UV} - y_{t,\rm IR} &= \frac{y_{t,\rm IR} y^2}{64 \pi^2}. 
\end{align}

\subsection{Singlet-doublet model RGE}\label{subsec:sdrge}

Since one of the new fermions is not a gauge singlet, it contributes to the running of couplings even when the new physics Yukawa couplings vanish. As such, we include the entire beta functions for this model. For the SM couplings, we find
\begin{align}
    \frac{\text{d}\lambda}{\text{d}\log \mu} = \frac{1}{(4 \pi)^2} \bigg[& 24 \lambda^2 -6 Y_4 +\frac{9 g^4}{8}+\frac{3 \left(g'\right)^4}{8}+\frac{3}{4} g^2 \left(g'\right)^2-2 y_1^4-2 y_2^4-4 y_1^2 y_2^2 \nonumber \\ 
    &+ \lambda  \left(12 Y_2 +4 y_1^2+4 y_2^2 -9 g^2-3 \left(g'\right)^2\right)\bigg] \nonumber \\ 
    + \frac{1}{(4 \pi)^4} \bigg[& 30 Y_6 +10 y_1^6 +10 y_2^6  +34 (y_1^4 y_2^2 + y_1^2 y_2^4 ) -6 (y_t^4 y_b^2 + y_t^2 y_b^4 ) -32 g_3^2 (y_t^4 +y_b^4) \nonumber\\ 
     &+(g^\prime)^2 \left(-\frac{8}{3} y_t^4 +\frac{4}{3} y_b^4 -4 y_{\tau }^4 \right)  + \frac{1}{4} \left(g'\right)^4 \left(-19 y_t^2 + 5 y_b^2 - 25 y_{\tau }^2  - y_1^2 - y_2^2 \right) \nonumber \\ 
     & - \frac{3}{4} g^4 \left(3 Y_2 + y_1^2 + y_2^2 \right) +\frac{1}{2} \left(g'\right)^2 g^2 \left( 21 y_t^2 +9 y_b^2  + 11 y_\tau^2-y_1^2 +  7 y_2^2  \right) \nonumber \\ 
     &+\frac{273 g^6}{16} -\frac{137 \left(g'\right)^6}{16} -\frac{107}{16} \left(g'\right)^2 g^4-\frac{197}{16} \left(g'\right)^4 g^2  \nonumber \\ 
     & + \lambda y_t^2 \left( \frac{45}{2} g^2 + \frac{85}{6} \left(g'\right)^2 +80 g_3^2 \right) + \lambda y_b^2 \left( \frac{45}{2} g^2 +\frac{25}{6} \left(g'\right)^2 +80 g_3^2  \right) \nonumber \\
     & + \lambda  y_{\tau }^2 \left( \frac{15}{2}  g^2 +\frac{25}{2}  \left(g'\right)^2\right)  +  \lambda ( y_1^2 + y_2^2) \left( \frac{15}{2}  g^2 +\frac{5}{2} \left(g'\right)^2 \right)  \nonumber \\
    &+\lambda  \left(-3 Y_4 - y_1^4-y_2^4 -\frac{33 g^4}{8} +\frac{39}{4} \left(g'\right)^2 g^2+\frac{223 \left(g'\right)^4}{8}+12 y_1^2 y_2^2-42 y_b^2 y_t^2\right) \nonumber \\ 
    &+\lambda^2 \left(108 g^2 +36 \left(g'\right)^2 -144 Y_2 -48 y_1^2-48 y_2^2 \right)-312 \lambda ^3\bigg], \\
    \frac{\text{d} y_t}{\text{d}\ln \mu} = \frac{y_t}{( 4 \pi)^2}\bigg[ &  \frac{9}{2} y_t^2 + \frac{3}{2} y_b^2 + y_{\tau }^2 +y_1^2 +y_2^2 -\frac{9 g^2}{4}-\frac{17}{12} \left(g'\right)^2 - 8 g_3^2 \bigg]  \nonumber \\
    +\frac{y_t}{(4 \pi)^4}\bigg[& 6 \lambda^2 -12 \lambda  y_t^2 -12 y_t^4 -\frac{1}{4} y_b^4 -\frac{9}{4} y_{\tau }^4 -\frac{9}{4} y_1^4 -\frac{9}{4} y_2^4 \nonumber \\
    &-\frac{11}{4} y_t^2 y_b^2 -\frac{9}{4} y_t^2 y_{\tau }^2 +\frac{5}{4} y_b^2 y_{\tau }^2  +\frac{5}{4} (y_1^2 + y_2^2 ) y_b^2 -\frac{9}{4} ( y_1^2 + y_2^2 ) y_t^2 - 5 y_1^2 y_2^2 \nonumber \\ 
    & + g^2 \left(\frac{225}{16}y_t^2+ \frac{99}{16} y_b^2 +\frac{15}{8} y_{\tau }^2   +\frac{15}{8} ( y_1^2 + y_2^2 )\right) \nonumber \\ 
    &+  \left(g'\right)^2 \left( \frac{131}{16} y_t^2 + \frac{7}{48} y_b^2  +  \frac{25}{8}  y_{\tau }^2+ \frac{5}{8} (y_1^2 +y_2^2 ) \right) +4 g_3^2 ( 9 y_t^2 + y_b^2 ) \nonumber \\
    &  -\frac{21 g^4 }{4}  +\frac{1303}{216} \left(g'\right)^4  -108 g_3^4 + 9 g^2 g_3^2+\frac{19}{9} g_3^2 \left(g'\right)^2 -\frac{3}{4} g^2 \left(g'\right)^2  \bigg] , \\
    \frac{\text{d} y_b}{\text{d}\ln \mu} = \frac{y_b}{(4 \pi)^2}\bigg[& \frac{3}{2} y_t^2 +\frac{9 y_b^2}{2} + y_{\tau }^2+y_1^2 +y_2^2  -\frac{9 g^2 }{4}-\frac{5}{12} \left(g'\right)^2-8 g_3^2  \bigg] \nonumber \\
    +\frac{y_b}{(4 \pi)^4}\bigg[& 6 \lambda^2  -12 \lambda  y_b^2 -\frac{1}{4}  y_t^4 -12 y_b^4 -\frac{9}{4}  y_{\tau }^4  -\frac{9}{4} y_1^4 -\frac{9}{4} y_2^4 \nonumber \\ 
    & -\frac{11}{4} y_b^2 y_t^2 +\frac{5}{4} y_t^2 y_{\tau }^2 +\frac{5}{4} (y_1^2 + y_2^2 ) y_t^2 -\frac{9}{4} y_b^2 y_{\tau }^2-\frac{9}{4} (y_1^2 + y_2^2 ) y_b^2  -5 y_1^2 y_2^2 \nonumber \\
    & +g^2 \left( \frac{99}{16} y_t^2  +\frac{225}{16} y_b^2 +\frac{15}{8} y_{\tau }^2 +\frac{15}{8} (y_1^2 + y_2^2) \right) \nonumber \\
    &  +(g^\prime)^2 \left( \frac{91}{48}  y_t^2 +\frac{79}{16} y_b^2 +\frac{25}{8}  y_{\tau }^2 +\frac{5}{8} ( y_1^2 +y_2^2 ) \right) +g_3^2 (4  y_t^2 + 36  y_b^2)  \nonumber \\
    & -\frac{21 g^4 }{4} -\frac{131}{216}  \left(g'\right)^4 -108 g_3^4  -\frac{9}{4} g^2  \left(g'\right)^2  +9 g^2 g_3^2  +\frac{31}{9} g_3^2  \left(g'\right)^2
    \bigg], \\
    \frac{\text{d} y_\tau}{\text{d}\ln \mu} = \frac{y_{\tau }}{(4\pi)^2}\bigg[& 3 y_t^2 + 3 y_b^2 +\frac{5 y_{\tau }^2}{2}  +y_1^2 +y_2^2 -\frac{9 g^2 }{4}-\frac{15}{4} \left(g'\right)^2 \bigg] \nonumber  \\
    +\frac{y_{\tau}}{(4 \pi)^4}\bigg[& 6 \lambda^2 -12 \lambda  y_{\tau }^2 -\frac{27}{4} y_t^4 -\frac{27}{4} y_b^4 -3 y_{\tau }^4 \nonumber \\ 
    -&\frac{9}{4} (y_1^2 + y_2^2) y_{\tau }^2-5 y_1^2 y_2^2 +\frac{3}{2} y_b^2 y_t^2  -\frac{27}{4} y_t^2 y_{\tau }^2 -\frac{27}{4} y_b^2 y_{\tau }^2 \nonumber \\
    &+g^2 \left( \frac{45}{8} y_t^2 + \frac{45}{8}  y_b^2  +\frac{165}{16} y_{\tau }^2 +\frac{15}{8} ( y_1^2 + y_2^2 ) \right) \nonumber \\
    &+ \left(g'\right)^2 \left( \frac{85}{24}  y_t^2 + \frac{25}{24} y_b^2  +\frac{179}{16}  y_{\tau }^2 +\frac{5}{8}  (y_1^2  +y_2^2 ) \right) +20 g_3^2 (y_t^2 + y_b^2 ) \nonumber \\
    &  -\frac{21 g^4 }{4} +\frac{167}{8} \left(g'\right)^4 +\frac{9}{4} g^2 \left(g'\right)^2 \bigg], \\
    \frac{\text{d}g^\prime}{\text{d}\ln \mu} = \frac{\left(g'\right)^3}{(4 \pi)^2}&\frac{15 }{2}+ \frac{\left(g'\right)^3}{(4 \pi)^4} \bigg[-\frac{17}{6}  y_t^2-\frac{5}{6} y_b^2 -\frac{5}{2}  y_{\tau }^2-\frac{1}{2} ( y_1^2 + y_2^2 ) +6 g^2 +\frac{104 \left(g'\right)^2}{9} +\frac{44}{3} g_3^2 \bigg] \nonumber \\
    + \frac{\left(g'\right)^3}{(4 \pi)^6} \bigg[ & -3 \lambda^2 + \frac{315}{16} y_t^4 + \frac{95}{16} y_b^4 +\frac{153}{16}  y_{\tau }^4 +\frac{45}{16} (y_1^4 + y_2^4 ) + \frac{123}{8} y_b^2  y_t^2 +\frac{199}{12} y_{\tau }^2  y_t^2 \nonumber \\ 
    & + \frac{157}{12} y_b^2 y_{\tau }^2 +\frac{73}{12} (y_1^2 +y_2^2 ) y_t^2 +\frac{31}{12} ( y_1^2 + y_2^2) y_b^2  +\frac{19}{4} (y_1^2 +y_2^2 ) y_{\tau }^2  +\frac{1}{2} y_1^2 y_2^2\nonumber \\ 
    &+ g^2 \left( \frac{3}{2} \lambda -\frac{785}{32} y_t^2 -\frac{437}{32} y_b^2 -\frac{543}{32}  y_{\tau }^2 -\frac{195}{32} (y_1^2 + y_2^2 )  \right) \nonumber \\
    &+ \left(g'\right)^2 \left(\frac{3}{2} \lambda -\frac{2827}{288}  y_t^2 -\frac{1267}{288} y_b^2 -\frac{281}{32} y_{\tau }^2-\frac{65}{32} (y_1^2 + y_2^2 ) \right)  \nonumber \\
    &+ g_3^2 \left( -\frac{29}{3} y_t^2 -\frac{17}{3}  y_b^2 \right) +\frac{1363}{64} g^4 -\frac{442429 \left(g'\right)^4}{5184} +99 g_3^4  \nonumber \\
    &+\frac{169}{96} g^2 \left(g'\right)^2-g^2 g_3^2  -\frac{137}{27} g_3^2 \left(g'\right)^2 \bigg], \\
    \frac{\text{d}g}{\text{d}\ln \mu} = \frac{g^3}{ (4 \pi)^2}\bigg[ & -\frac{5}{2}\bigg] + \frac{g^3}{(4 \pi)^4} \bigg[-\frac{1}{2} Y_2 -\frac{1}{2} (y_1^2 + y_2^2 ) +14 g^2 +2 \left(g'\right)^2  +12 g_3^2 \bigg] \nonumber \\
    +\frac{g^3}{(4\pi)^6}\bigg[&-3 \lambda^2 + \frac{147}{16} (y_t^4 + y_b^4 )+\frac{29}{16} y_{\tau }^4 +\frac{29}{16} (y_1^4 + y_2^4)  +\frac{5}{2} y_1^2 y_2^2 +\frac{117}{8} y_b^2 y_t^2 \nonumber \\ 
    & +\frac{15}{4} y_{\tau }^2 y_t^2  +\frac{15}{4} y_b^2 y_{\tau }^2 +\frac{15}{4} (y_1^2 + y_2^2) y_t^2 +\frac{15}{4} (y_1^2 + y_2^2) y_b^2   +\frac{5}{4} (y_1^2 + y_2^2) y_{\tau }^2  \nonumber \\ 
    & + g^2 \left(\frac{3 \lambda }{2} -\frac{729}{32} y_t^2 -\frac{729}{32} y_b^2 -\frac{243}{32} y_{\tau }^2 -\frac{243}{32} (y_1^2+ y_2^2 ) \right)\nonumber \\ 
    &+ \left(g'\right)^2 \left(\frac{1}{2} \lambda  -\frac{533}{96} y_b^2  -\frac{593}{96}  y_t^2 - \frac{85}{32}  y_{\tau }^2 -\frac{65}{32} ( y_1^2  + y_2^2 ) \right) -7 g_3^2 (y_t^2 + y_b^2 ) \nonumber \\ 
    &+\frac{49313 g^4}{192} -\frac{8213}{576} \left(g'\right)^4 + 81 g_3^4  +\frac{343}{32} g^2 \left(g'\right)^2 +39 g_3^2 g^2 -\frac{1}{3} g_3^2 \left(g'\right)^2\bigg],\\
    \frac{\text{d}g_3}{\text{d}\ln \mu} = \frac{g_3^3}{(4 \pi)^2}\big[ & -7\big] + \frac{g_3^3}{(4 \pi)^4}\bigg[-2 y_t^2 -2 y_b^2+\frac{9}{2} g^2 +\frac{11}{6}  \left(g^\prime \right)^2-26 g_3^2 \bigg] \nonumber \\
    +\frac{g_3^3}{(4 \pi)^6}\bigg[& 15 (y_t^4 +  y_b^4) +18 y_b^2 y_t^2 +\frac{7}{2} (y_t^2 + y_b^2 ) y_{\tau }^2 +\frac{7}{2} (y_1^2 + y_2^2) (y_b^2 + y_t^2 ) \nonumber \\
    &-\frac{93}{8} g^2 (y_t^2 + y_b^2 ) -\left(g'\right)^2 \left( \frac{101}{24} y_t^2  +\frac{89}{24}  y_b^2  \right) -40 g_3^2 (y_t^2 + y_b^2 ) \nonumber \\
    & +\frac{87}{8} g^4 -\frac{2857}{216} \left(g'\right)^4 +\frac{65 g_3^4}{2} -\frac{1}{8} g^2  \left(g'\right)^2  +21 g^2 g_3^2 +\frac{77}{9} g_3^2 \left(g'\right)^2 \bigg].
\end{align}
For the two new Yukawa couplings meanwhile, we find
\begin{align}
    \frac{\text{d}y_1}{\text{d}\ln \mu} = \frac{y_1}{(4 \pi)^2}\bigg[&3 Y_2+ \frac{5 y_1^2}{2}+4 y_2^2 -\frac{9 g^2 }{4}-\frac{3}{4}  \left(g'\right)^2 \bigg] \nonumber \\
    +\frac{y_1}{(4 \pi)^4}\bigg[& 6 \lambda^2 -12 \lambda  (y_1^2 + y_2^2 ) -\frac{27}{4} y_t^4 -\frac{27}{4} y_b^4 -\frac{9}{4} y_{\tau }^4 -3 y_1^4 -9 y_2^4\nonumber \\ 
    & +\frac{3}{2}  y_b^2 y_t^2-\frac{27}{4} (y_1^2 + y_2^2 ) y_b^2 -\frac{27}{4} y_1^2 y_t^2-\frac{21}{2} y_2^2 y_t^2  -\frac{9}{4} y_1^2 y_{\tau }^2 -\frac{7}{2} y_2^2  y_{\tau }^2-15 y_2^2 y_1^2 \nonumber \\
    &+ g^2 \left(\frac{45}{8}  y_t^2 +\frac{45}{8} y_b^2 +\frac{15}{8} y_{\tau }^2 +\frac{165}{16} y_1^2+\frac{39}{4} y_2^2 \right) \nonumber \\
    & + (g^\prime)^2 \left(\frac{85}{24}  y_t^2 +\frac{25}{24}  y_b^2  +\frac{25}{8} y_{\tau }^2 +\frac{103}{16} y_1^2 +\frac{1}{4} y_2^2 \right) +20 g_3^2 (y_t^2 + y_b^2 ) \nonumber \\
    &-\frac{21 g^4 }{4} +\frac{13}{8} \left(g'\right)^4 -\frac{9}{4} g^2 \left(g'\right)^2 \bigg], \\
    \frac{\text{d}y_2}{\text{d}\ln \mu} = \frac{y_2}{(4 \pi)^2}\bigg[&3 Y_2 +4 y_1^2 +\frac{5 y_2^2}{2} -\frac{9 g^2}{4}-\frac{3}{4} \left(g'\right)^2 \bigg] \nonumber \\
    +\frac{y_2}{(4 \pi)^4}\bigg[& 6 \lambda^2 -12 \lambda ( y_1^2+ y_2^2) -\frac{27}{4} y_t^4 -\frac{27}{4} y_b^4 -\frac{9}{4} y_{\tau }^4 - 9 y_1^4 -3 y_2^4 \nonumber \\
    &+\frac{3}{2} y_b^2 y_t^2 -\frac{21}{2} y_1^2 y_t^2  -\frac{27}{4} y_2^2 y_t^2 -\frac{21}{2} y_1^2 y_b^2 -\frac{27}{4} y_2^2 y_b^2 -\frac{7}{2} y_1^2 y_{\tau }^2 -\frac{9}{4} y_2^2 y_{\tau }^2 -15 y_1^2 y_2^2  \nonumber \\
    &+ g^2 \left( \frac{45}{8} y_t^2+ \frac{45}{8}  y_b^2 +\frac{15}{8}  y_{\tau }^2 +\frac{39}{4} y_1^2 +\frac{165}{16} y_2^2 \right) \nonumber \\
    &+ (g^\prime)^2 \left(\frac{85}{24}  y_t^2 +\frac{25}{24} y_b^2 + \frac{25}{8} y_{\tau}^2 +\frac{1}{4} y_1^2 +\frac{103}{16} y_2^2 \right) +20 g_3^2 (y_t^2 + y_b^2) \nonumber \\
    &-\frac{21 g^4}{4} +\frac{13}{8} \left(g'\right)^4-\frac{9}{4} g^2 \left(g'\right)^2 \bigg].
\end{align}
The form of the threshold correction for this model is more complicated than for the heavy neutral lepton model due to the presence of two scales, $m_S$ and $m_D$. We first perform the matching at the larger of the two scales, $m_D$. The threshold corrections then take the form
\begin{align}
    \lambda_{\rm UV} - \lambda_{\rm IR}&=\frac{1}{16 \pi ^2 \left(m_D^2-m_S^2\right){}^3}\bigg[4 y_1 y_2 m_D^5 m_S \left(2 \left(y_1^2+y_2^2\right)-\lambda _{\text{IR}}\right)  \nonumber\\
    &+4 y_1 y_2 m_D m_S^5 \left(\lambda _{\text{IR}}-2 \left(y_1^2+y_2^2\right)\right)  \nonumber \\
    &+m_D^4 m_S^2 \left(-12 y_2^2 y_1^2 \log \left(\frac{m_D^2}{m_S^2}\right)-7 \left(y_1^2+y_2^2\right) \lambda _{\text{IR}}+3 y_1^4+14 y_2^2 y_1^2+3 y_2^4\right) \nonumber \\
    &-8 y_1 y_2 m_D^3 m_S^3 \left(2 \left(y_1^2+y_2^2\right)-\lambda _{\text{IR}}\right) \log \left(\frac{m_D^2}{m_S^2}\right) \nonumber \\
    &-m_D^2 m_S^4 (-\left(y_1^2+y_2^2\right) \lambda _{\text{IR}} \left(6 \log \left(\frac{m_D^2}{m_S^2}\right)+7\right)+\left(5 y_1^4+14 y_2^2 y_1^2+5 y_2^4\right) \log \left(\frac{m_D^2}{m_S^2}\right) \nonumber \\
    &+2 \left(y_1^4+8 y_2^2 y_1^2+y_2^4\right)) \nonumber \\
    &+\left(y_1^2+y_2^2\right) m_S^6 \left(\left(y_1^2+y_2^2\right) \left(\log \left(\frac{m_D^2}{m_S^2}\right)-1\right)-\lambda _{\text{IR}} \left(2 \log \left(\frac{m_D^2}{m_S^2}\right)+1\right)\right)   \nonumber\\
    &+m_D^6 \left(\left(y_1^2+y_2^2\right) \lambda _{\text{IR}}+4 y_1^2 y_2^2\right)) \bigg], \\
    y_{t,\rm UV} - y_{t,\rm IR} &= -\frac{y_t}{64 \pi ^2 \left(m_D^2-m_S^2\right)^3}\bigg[-4 y_1 y_2 m_D^5 m_S-7 \left(y_1^2+y_2^2\right) m_D^4 m_S^2+4 y_1 y_2 m_D m_S^5  \nonumber \\
    &+8 y_1 y_2 m_D^3 m_S^3 \log \left(\frac{m_D^2}{m_S^2}\right)+\left(y_1^2+y_2^2\right) m_D^2 m_S^4 \left(6 \log \left(\frac{m_D^2}{m_S^2}\right)+7\right)  \nonumber \\
    &-\left(y_1^2+y_2^2\right) m_S^6 \left(2 \log \left(\frac{m_D^2}{m_S^2}\right)+1\right)+\left(y_1^2+y_2^2\right) m_D^6\bigg]. 
\end{align}
For completeness, we also provide the threshold corrections for a matching at $m_S$:
\begin{align}
\lambda_{\text{UV}}- \lambda_{\text{IR}}
=&
\frac{1}{16 \pi^2 \left(m_D^2 - m_S^2\right)^3}
\bigg[4 m_D^5  m_S   y_1  y_2 \left(2  (y_1^2 + y_2^2) - \lambda_{\rm IR} \right)\nonumber \\
&+4  m_D  m_S^5 y_1  y_2 \left(\lambda_{\rm IR} - 2  (y_1^2 + y_2^2) \right) \nonumber \\ 
&+m_D^4  m_S^2  \left(-12  y_1^2  y_2^2  \log\left( \frac{m_D^2}{m_S^2} \right) - 
   7  (y_1^2 + y_2^2 ) \lambda_{\rm IR} + 3  y_1^4 + 14   y_1^2  y_2^2 + 
   3   y_2^4 \right. \nonumber \\ 
   &\left. - 
   3   (y_1^2 + y_2^2) ((y_1^2 + y_2^2) - 2 \lambda_{\rm IR})  \log\left( \frac{m_D^2}{m_S^2} \right) \right) \nonumber \\
&-8 m_D^3  m_S^3 y_1  y_2  (2  (y_1^2 + 2  y_2^2 ) - \lambda_{\rm IR}) \log \left( \frac{m_D^2}{m_S^2} \right) \nonumber \\ 
&-m_D^2  m_S^4  (2  y_1^4 + 16   y_1^2  y_2^2 + 2  y_2^4 + 
   2 (y_1^4 + 4 y_1^2  y_2^2 + y_2^4 ) \log\left( \frac{m_D^2}{m_S^2} \right) - 
   7 (y_1^2 + y_2^2 ) \lambda_{\rm IR})\nonumber \\
&-m_S^6 (y_1^2 + y_2^2 )  (y_1^2 + y_2^2 + \lambda_{\rm IR})\nonumber  \\
&+m_D^6  (4  y_1^2  y_2^2 + (y_1^2 + y_2^2) \lambda_{\rm IR} + (y_1^2 + 
      y_2^2) ((y_1^2 + y_2^2 ) - 2  \lambda_{\rm IR}) \log\left( \frac{m_D^2}{m_S^2} \right) )\bigg] \, , \\ 
y_{t,\rm UV} - y_{t,\rm IR} &= -\frac{y_t}{64 \pi ^2 \left(m_D^2-m_S^2\right)^3}\bigg[m_D^6 (y_1^2 + y_2^2)
-4 m_D^5 m_S y_1 y_2 -7 m_D^4 m_S^2 (y_1^2 + y_2^2)\nonumber \\
&+7 m_D^2 m_S^4 (y_1^2 + y_2^2) +4 m_D m_S^5 y_1 y_2 -m_S^6 y_1^2 - m_S^6 y_2^2\nonumber \\
& +2 (3  m_D^4  m_S^2  (y_1^2 + y_2^2) + 4  m_D^3  m_S^3  y_1  y_2 - 
   m_D^6  (y_1^2 + y_2^2)) \log\left(\frac{m_D^2}{m_S^2} \right) \bigg]
\end{align}
 
\bibliographystyle{JHEP}
\bibliography{references}

\end{document}